# CAN NETWORK THEORY-BASED TARGETING INCREASE TECHNOLOGY ADOPTION?

By

Lori Beaman, Ariel BenYishay, Jeremy Magruder, and Ahmed Mushfiq Mobarak

# Can Network Theory-based Targeting Increase Technology Adoption?

| Lori Beaman | Ariel BenYishay | Jeremy Magruder | Ahmed Mushfiq Mobarak |
|---|---|---|---|
| Northwestern Univ. | Coll. of William and Mary | UC-Berkeley | Yale University |

August 2018

Abstract

In order to induce farmers to adopt a productive new agricultural technology, we apply simple and complex contagion diffusion models on rich social network data from 200 villages in Malawi to identify seed farmers to target and train on the new technology. A randomized controlled trial compares these theory-driven network targeting approaches to simpler strategies that either rely on a government extension worker or an easily measurable proxy for the social network (geographic distance between households) to identify seed farmers. Our results indicate that technology diffusion is characterized by a complex contagion learning environment in which most farmers need to learn from multiple people before they adopt themselves. Network theory based targeting can out-perform traditional approaches to extension, and we identify methods to realize these gains at low cost to policymakers.

* Contact: Beaman: l-beaman@northwestern.edu, BenYishay: abenyishay@wm.edu, Magruder: jmagruder@berkeley.edu, Mobarak: ahmed.mobarak@yale.edu. We thank the CEGA/JPAL Agricultural Technology Adoption Initiative (ATAI) and 3ie for financial support. Beaman acknowledges support by the National Science Foundation under grant no. 1254380. We gratefully acknowledge the support and cooperation of Paul Fatch, Readwell Musopole and many other staff members of the Malawi Ministry of Agriculture. Thomas Coen, Niall Kelleher, Maria Jones, Ofer Cohen, Allen Baumgardner-Zuzik, and the IPA-Malawi country office provided invaluable support for data collection. Hossein Alidaee and Tetyana Zelenska provided excellent research assistance. We thank, without implicating, Arun Chandrasekhar, Matt Jackson, Kaivan Munshi, Chris Udry and numerous seminar audiences for very helpful comments.

## 1. Introduction

Technology diffusion is critical for growth and development (Alvarez et al. 2013, Perla and Tonetti 2014). Information frictions are potential constraints to technology adoption, and social relationships can serve as important vectors through which individuals learn about, and are then convinced to adopt, new technologies.[1] With a better understanding of the diffusion process and how people choose to adopt new technologies, we could potentially manipulate social learning and identify strategies that would maximize diffusion. In this paper we specify a model of learning, pair it with a field experiment in which we choose "optimal" entry points according to that theory of network diffusion, and introduce a productive new agricultural technology via those entry points across 200 villages in Malawi. Our goal is to test whether insights from network theory can be practically useful in enhancing the diffusion of a technology in the field, and in the process, generate new evidence on the nature of social learning and diffusion.

Our experiment is on agricultural extension because a large literature has established that social learning plays a central role in diffusing agricultural technologies (Griliches 1957, Foster and Rosenzweig 1995, Munshi 2004, Bandiera and Rasul 2006, Conley and Udry 2010). Network theory is therefore particularly likely to be of practical value in that sector. The setting itself is also very important for development: 64.5% of the world's population living in poverty are engaged in agriculture (Castaneda et al 2016), and agricultural yields are especially low and slow-growing in Africa (World Bank 2008). Extension services are an important policy tool to counter low productivity: Developing country governments employ over 400,000 extension workers, and Anderson and Feder (2007) note that this "may well be the largest institutional development effort the world has ever known." The specific technology we promote, 'pit planting', has the potential to significantly improve maize yields in arid areas of rural Africa.[2] It is a practice that was largely unknown in Malawi, and learning is therefore crucial for the diffusion of this technology.

[1] Large literatures in economics (Munshi 2008, Duflo and Saez 2003, Magruder 2010, Beaman 2012), finance (Beshears et al. 2013, Bursztyn et al. 2013), sociology (Rogers 1962), and medicine and public health (Coleman et al 1957; Doumit et al 2007) show that information and behaviors spread through inter-personal ties.

We use the "threshold model" of diffusion (e.g. Granovetter 1978; Centola and Macy 2007; Acemoglu *et al* 2011) which postulates that individuals adopt a behavior only if they are connected to at least a threshold number of adopters, as the basis for our exploration. This is an important class of models for policy analysis, because under this particular framework, the choice of network entry points used to influence diffusion becomes crucial (Akbarpour, Malladi and Saberi 2017). We first develop a micro-foundation for this model that makes use of one central insight about the process of learning: If the technology is new and gathering information about it is costly, then farmers may rationally choose not to seek information when their network members do not know much about this technology. When net benefits of adoption are large and clear, then a single source of information may suffice to encourage learning (and then, adoption), and the "threshold" is very low. Technology diffusion then looks like viral infection, and this process is referred to as "simple contagion". When learning is difficult because each connection cannot provide accurate, relevant information, then farmers may need data from multiple sources, and the threshold increases. The resulting learning process is called "complex contagion".

Threshold models yield specific predictions on network entry points that are likely to be most effective, in theory, to quickly diffuse a new technology. Suppose an extension agency can train two farmers in a new technology. Under simple contagion, if the extension agency wants to maximize adoption over the next 3-5 years, it would spread the entry points far apart to minimize repetition and redundancy in the same part of the network. Instead, if technology diffusion follows complex contagion, it is critical that the trained farmers are clustered together and share connections, in order to improve the chances that some recipients will learn from multiple sources simultaneously.[3] The design of our field experiment is based on such insights.

---

[2] It has been shown to increase productivity by 50-100% in tests conducted under controlled conditions (Haggblade and Tembo 2003); in large-sample field tests conducted under realistic "as implemented by government" conditions (BenYishay and Mobarak 2018), and using experimental variation among villagers in this study.

Before implementing the experiment, we collect social network census data on agricultural learning relationships in 200 villages in Malawi. We then conduct simulations on those data to identify the theoretically optimal entry points ("seeds") that would maximize diffusion of information about a new technology, assuming the diffusion process is characterized by either simple contagion or complex contagion. Villages are then randomly assigned one of those targeting strategies (either simple or complex contagion), and the Malawi extension services trained the seeds we selected via the theoretical simulations. Seeds were asked after training to disseminate the information. We then trace adoption patterns in these villages over the next 2-3 years.

We compare the adoption in these network theory-based treatment villages against a benchmark treatment applied to set of other randomly selected villages, where agricultural extension agents use local knowledge to select seeds. Typically, this involves asking village leaders to nominate a pair of extension partners and is similar to what many extension workers normally do outside of our study context.[4] As another comparison, we implement a fourth treatment in which we select optimal seeds who would maximize diffusion under complex contagion but assuming geographic proximity proxies for social network connections. Unlike social network relationships, geographic location is easy for extension agents to observe, so we view this as a first step towards a policy-relevant alternative to the data intensive network-theory based approaches.

---

[3] The need to cluster trained farmers under complex contagion implies that the diffusion gains from strategically targeting seeds cannot be easily replicated by simply training a few additional farmers without worrying about targeting (Akbarpour *et al* 2018, Jackson and Storms 2018). Akbarpour *et al* 2018 show in many other probabilistic diffusion models where a single connection is sufficient for diffusion (including the model estimated in Banerjee et al 2013), targeting is not that important because randomly choosing just a few additional seeds outperforms (both in theory and in simulations) the strategy of carefully targeting seeds based on their network position. By contrast, Banerjee, Breza, Chandrasekhar and Golub (2018) show that information broadcasting can generate less information flow than seeding, when individuals need to ask follow up questions about information received within a network.

[4] Extension workers may be able to select influential partners based on specialized knowledge such as her eagerness to try the new technology, or the trust other villagers place in their opinions. As such, this benchmark provides a demanding test for network-based diffusion theory: our theoretically optimal partners were selected only by their position in the network, without the advantage of this additional local information.

We find that the theory-driven targeting of optimal seed farmers leads to greater technology diffusion than the benchmark approach. Threshold theory-based targeting increases adoption by 3 percentage points more than relying on extension workers to choose seeds, during the 3-year period of the experiment when pit planting adoption grew from 0% to about 10% in these villages overall.[5] We use our micro data on exactly which farmers adopt to provide more direct evidence in favor of the learning model we postulate. For example, we document larger and more sustained gains in adoption from the network-theory based targeting for the subset of farmers for whom returns to this technology are high (given their land-type), and in villages where farmers were initially uninformed about the technology, as predicted by theory.

The complex contagion model suggests that one of the potential consequences of poor targeting is complete failure to adopt within the village. We observe no diffusion of pit planting in 45% of the 'benchmark' villages after 3 years. In villages where seeds were selected using the complex contagion model, there was a 56% greater likelihood that at least one person other than the seeds adopts in the village, relative to the benchmark. The results suggest that simply changing *who* is trained in a village on a technology on the basis of social network theory can increase the adoption of new technologies compared to the Ministry's existing extension strategy.

Even the low-cost geography-based targeting strategy generates some gains in adoption relative to the benchmark. However, physical proximity does not appear to be a good proxy for social connections in this context. Developing other low-cost proxies for social network structure would be a useful avenue for future research.[6] As a first step, we develop an intuitive algorithm to identify productive extension partners that can be implemented with a small number of interviews, and simulations on our data show that this method would generate large gains in technology adoption.[7]

[5] This rate of increase in adoption is not unusual for new agricultural technologies, including very profitable ones (e.g. Munshi 2007). Ryan and Gross (1943) show that it took 10 years for hybrid seed corn to be adopted in Iowa in the 1930s, and there was often 5 years between when a farmer heard of the technology and adopted it.

The existing literature has shown that more extensive diffusion takes place when entry points are more central (Banerjee et al 2013 in the context of microfinance in India; Kim et al 2015 looking at health behaviors in Honduras). Our exercise provides insights into engineering more extensive diffusion in a context of social learning, and also helps us examine whether network theory has predictive power[8] to speed up the diffusion of a policy-relevant technology.

The rest of the paper is organized as follows. We present the theoretical model on which the experimental design is based in Section 2. Section 3 describes the experimental setting and design. Section 4 discusses all field activities, including the intervention and data collection. Section 5 describes the characteristics and activities of the seed farmers and the performance of the technology in the field and documents that social diffusion took place. Section 6 presents the village (or network) level experimental results. Section 7 uses simulations on our data to explore cost-effective, policy-relevant alternatives to the data-intensive network-theory based procedures we experiment with in this paper. Section 8 concludes.

---

[6] For example, promising results in Banerjee et al (2018) imply that households know who is central in their village, and this type of information may be easily elicited from a random sample of people. Kim et al (2015) use a related elicitation mechanism based on friends-of-friends, and conduct an experiment to distribute public health coupons in a sample of 32 villages.

[7] A variety of other papers test the ability of local institutions, such as nominations or focus groups, to identify useful partners: Kremer et al (2011) identify and recruit 'ambassadors' to promote water chlorination in rural Kenya, Miller and Mobarak (2014) first markets improved cookstoves to 'opinion leaders' in Bangladeshi villages before marketing to others, and BenYishay and Mobarak (2015) incentivize 'lead farmers' and 'peer farmers' to partner with agricultural extension officers in Malawi.

[8] In contrast, Carrel et al (2013) offers a cautionary tale that data-driven attempts to manipulate social interactions do not have predictive power to design optimal classrooms.

## 2. Theoretical Model Motivating the Experimental Design

### 2.1 A Micro-Foundation for the Threshold Model of Diffusion

The linear threshold model (Granvetter 1978; Acemoglu *et al* 2011) is one of the seminal descriptions of diffusion processes. This model posits that an agent will adopt a new behavior once at least $\lambda$ of his connections adopt the technology. We base our experimental design on this class of models for three reasons. First, the threshold model is built on a very natural insight about how social learning might affect adoption decisions: A farmer learns from the behavior of each connection she has, and depending on the depth of the farmer's priors, it may take more or fewer connections to motivate her to change behavior. The threshold formulation is therefore more naturally micro-founded with a model of learning, relative to other canonical diffusion models.[9] Second, the threshold model has served as an important building block for diffusion theory. The original paper that introduced this formulation (Granovetter 1978) has been cited about 5000 times on google scholar. Third, the formulation is consistent with some key empirical patterns about technology diffusion in agriculture. For example, the number of contacts acting as a key driver of adoption decisions can explain the well-known S-shaped diffusion pattern for new technologies (e.g. Griliches 1957).[10]

Perhaps due to the interdisciplinary interest in the threshold model, there is little consensus on the mechanisms underlying the threshold model or associated empirical predictions. This section therefore formally derives the threshold model as the outcome of optimizing behavior of microeconomic agents, so that we can take some clear predictions to the experiment and the micro data. We develop this micro-foundation by extending a framework presented in Banerjee *et al* (2016) (hence: BBCM). One key insight in BBCM is that the majority of members of a social network may not have access to *any* useful signal when they are confronted with an entirely new technology. Thus, there are two parts to the learning problem for new technologies: acquiring a signal in the first place (becoming informed) which may be costly, and forming a revised belief on the profitability of the new technology based on the signals received from informed connections. Optimizing farmers adopt a new technology only if their beliefs change and they are convinced by others that this would be more profitable than alternatives.[11]

[9] Both simple and complex contagion formulations are related to Bayesian learning models. In simple contagion, a single contact motivates adoption, suggesting that a person's prior (to not adopt) is not very strong. In contrast, complex contagion suggests that additional observations of adoption are necessary to move most people's priors. We will use this simpler version rather than a formal Bayesian learning model as those models quickly become intractable in real world networks (Chandrasekhar et al 2012).

[10] The slow rate of diffusion in early stages can be explained by not many people in a network having multiple contacts who have adopted when a technology is new, but the probability of having multiple adopter contacts increases more rapidly as the technology spreads through the network. In general, both this intuition and examples of threshold modeling have been unspecific as to whether the threshold is in the number of contacts, or the fraction of contacts. The micro-foundation we develop below produces a threshold in number of contacts.

There are three key phases of decision-making in our model: (1) the farmer has to decide whether to acquire information, (2) she has to combine the new information with her priors, and (3) she then decides whether to adopt the new technology. We will present and solve the model backwards, starting with the third phase.

The farmer will choose to adopt the new technology in phase 3 if she believes that adoption will be profitable. Suppose farmer *j* knows the technology will cost her $c_j$ to adopt and believes the new technology has either profit $\bar{\pi}$ or $\underline{\pi}$ $(\underline{\pi} < c_j < \bar{\pi})$.[12] Since the technology is new and farmer *j* is initially uninformed, she has a uniform prior as to whether the technology is profitable or not. She can aggregate signals given by her connections to update her prior and make an informed adoption decision.

[11] A very different micro-foundation for a similar model is explored in Jackson and Storms (2018). In that model, thresholds become relevant as individuals face greater payoffs from conforming to the behavior of their connections. Since coordination incentives for smallholder adoption of new agricultural technologies adoption seem likely to be low we pursue instead a model based on learning and individual optimization.

[12] Here for simplicity we follow BBCM in assuming that the distribution of profits is binary and known. In practice, there will be uncertainty over a wider range of profits due to the potential performance of the technology under different agroclimatic conditions and different weather realizations. While posterior distributions will be much more complicated under more realistic depictions of uncertainty, the key intuition driving the threshold model will be unchanged.

We adopt the same learning environment modeled in BBCM: First, informed farmer $i$ disseminates a binary signal, $x_i \in \{\underline{\pi}, \bar{\pi}\}$, which is accurate with probability $\alpha > \frac{1}{2}$. Uninformed farmers do not disseminate a signal. Second, farmers follow DeGroot learning (DeMarzo et al 2003). DeGroot learning can be interpreted as a boundedly rational version of Bayes learning, and suggests that farmers aggregate signals from their connections without attempting to calculate the inherent correlation structure between those signals (so if farmer *j* sees a signal of $\bar{\pi}$ from both farmers *i* and *k*, she interprets that as two positive signals without decomposing the likelihood that farmer *i* and *k* are disseminating information obtained from the same source).[13] Once farmers have observed signals from their informed connections, they aggregate those signals via Bayes' rule.

This framework suggests the following for the second phase of the farmer's learning problem: Suppose farmer *j* has $D_j$ informed contacts. If farmer *j* decides to learn about the new technology from her informed contacts, and if *H* of those contacts provide the signal $x = \bar{\pi}$, then the farmer's posterior probability that $\pi = \bar{\pi}$ is given by[14]

$$E_j[\pi = \bar{\pi}] = \frac{\alpha^{2H-D_j}}{\alpha^{2H-D_j} + (1-\alpha)^{2H-D_j}}$$

Denote $\tilde{\bar{\pi}} = \bar{\pi} - \underline{\pi}$ and $\tilde{c_j} = c_j - \underline{\pi}$. With that posterior, the farmer would adopt the technology if

$$\frac{\tilde{c}_j}{\tilde{\bar{\pi}}} \leq \frac{\alpha^{2H-D_j}}{\alpha^{2H-D_j}+(1-\alpha)^{2H-D_j}} \leq \frac{\alpha^{D_j}}{\alpha^{D_j}+(1-\alpha)^{D_j}} \tag{1}$$

This model highlights a potential challenge to diffusing new technologies: when few other farmers are informed, then there is a ceiling on how much a new farmer's priors would move even if they receive unanimously positive signals from the informed. At early stages in the diffusion process, $D_j$ may be small for many farmers.

[13] Chandrasekhar *et al* (2016) provide laboratory evidence in support of DeGroot learning over Bayes learning in India. Additional citations in favor of this boundedly-rational approximation can be found in BBCM.

[14] A simple proof is given in BBCM.

Last, we consider the first phase of the farmer's learning problem, which is her decision to acquire signals and become informed. Here, we depart from BBCM to suggest that there may be a small cost to receiving a signal $\eta$. This cost could be interpreted as shoe-leather costs of acquiring information (which are not necessarily trivial in villages in rural Malawi as households may be fairly far apart), or as stigma from seeking information (e.g. Breza and Chandrasekhar 2017).

Thus the farmer j with informed degree $D_j$ has an objective given by

$$\max_{d \leq D_j} \sum_{h \leq d} \frac{1}{2} \left( \left[ \alpha^{h}(1-\alpha)^{d-h}(\overline{\pi} - c_j) + (1-\alpha)^{h} \alpha^{d-h}(\underline{\pi} - c_j) \right] \left( I \left( \frac{\alpha^{2h-d}}{\alpha^{2h-d} + (1-\alpha^{2h-d})} > \frac{\tilde{c}}{\approx{\pi}} \right) \right) \right) - \eta d$$

When $\eta = 0$, the dynamics of learning are explored by BBCM. However, when $\eta > 0$ the dynamics are slightly different. In that case (for small $\eta$) farmers will only become informed if

$$\frac{\alpha^{D_j}}{\alpha^{D_j} + (1-\alpha)^{D_j}} > \frac{\tilde{c}_j}{\approx{\pi}} \tag{2}$$

In other words, farmers only choose to seek information if they have a large enough number of informed connections, such that it is possible that an informed decision would lead them to adopt. In this case (and for small $\eta$), farmers will choose to seek information when they have only one informed connection if

$$\frac{\alpha}{\alpha + (1-\alpha)} > \frac{\tilde{c}_j}{\approx{\pi}} \tag{3}$$

In general, they will choose to become informed with $\lambda$ informed connections if

$$\frac{\alpha^{\lambda}}{\alpha^{\lambda}+(1-\alpha)^{\lambda}} > \frac{\tilde{c}_j}{\tilde{\tilde{\pi}}} \tag{4}$$

This implies that farmers choose to become informed about new technologies if expectations about the net benefits of technology are high (i.e., low costs and high potential gains), or if signals from individual other farmers are highly accurate. Under certain parameter values, just a single informed contact may be sufficient to induce farmers to seek information. That is the diffusion process that Centola and Macy (2007) refer to as a "simple contagion". They demonstrate that some types of information – for example, job opportunities – spread in this way. On the other hand, if the expected upside of the technology is more modest relative to costs, or if signals from other farmers have low accuracy, then farmers may only be persuaded to seek information when there is sufficient information to be gained from their network.[15] In that case, for many farmers the lowest $\lambda$ satisfying equation (4) may be larger than 1, and information diffusion follows a process termed "complex contagion" in the literature.[16]

Our interpretation of the microeconomics of the threshold theory is that the thresholds result from an underlying process of farmers deciding *whether* to learn, given their information environment. This motivates an experimental design in which we seed new information in a network to improve the information environment, which should jump-start the technology diffusion process.

Given that the econometrician is unlikely to observe signal accuracy $(\alpha)$, the threshold required for adoption of a specific new technology is an empirical question. As a numerical example, consider a technology with 30% potential returns (so that $\tilde{\tilde{\pi}} = 1.3\ \tilde{c_J}$). If signals are more than 77% accurate, farmers will choose to become informed if they have a single informed connection, and diffusion will follow a simple contagion. If signal accuracy falls in the range 65% - 77% accurate, then farmers will only become informed if they have 2 informed connections, and learning will follow a complex contagion. If signals are less than 65% accurate, then farmers will need at least 3 informed connections to make an adoption decision.

---

[15] Though not explicitly considered here, minimal thresholds for learning will also be higher if $\eta$ (the cost of information acquisition) is larger.

[16] Several theory papers have explored the implications of this model. In contrast to the "strength of weak ties" in labor markets proposed by Granovetter (1978), strong ties may be important for the diffusion of behaviors that require reinforcement from multiple peers. Centola (2010) provides experimental evidence that health behaviors diffuse more quickly through networks where links are clustered, consistent with complex contagion. Acemoglu et al (2011) highlights that when contagion is complex, highly clustered communities will need a seed placed in the community in order to induce adoption. Finally, Monsted *et al* provide experimental evidence generated by twitter-bots that twitter hashtag retweets follow a process which more closely resembles complex than simple contagion.

## 2.2 Model Predictions and Implications for the Experiment

The micro-foundation of the threshold model suggests a particular structure for an experiment and its analysis. This model would need to be tested using the diffusion of a truly new technology, where would-be adopters are *ex ante* uninformed about the technology and face an important adoption decision. A corollary is that the threshold model should fit the data better in locations where the technology is more novel.

If thresholds exist and are above one, then seeding the network with multiple sources of information who are clustered in the same part of the network will achieve very different diffusion patterns than seeding the network with the same number of information sources spread more diffusely. Our experimental design will take advantage of this insight. The information environment only induces learning when initial nodes happen to share some connections, which is something we can test using micro data on technology diffusion patterns.

The model highlights that farmers will become informed when they have sufficiently many informed contacts. However, conditional on being informed, they will only adopt the technology if the realization of signals from their connections are sufficiently positive. These two facts grant two different tests of the model.

PREDICTION 1: If most farmers in a village have a threshold $\bar{\lambda}$, then people who are connected to at least $\bar{\lambda}$ informed farmers should become informed themselves.

PREDICTION 2: Adoption should increase most strongly among farmers who have high net benefits of adoption, who would adopt with a broader range of received signals.[17]

## 3. Field Experiment

### 3.1 Setting

Our experiment on technology diffusion via an agricultural extension system takes place in 200 villages randomly sampled from three Malawian districts with largely semi-arid climates (Machinga, Mwanza, and Nkhotakota). Approximately 80% of Malawi's population lives in rural areas (World Bank 2011), and agricultural production in these areas is dominated by maize: 97% of farmers grow maize, and over half of households grow no other crop (Lea and Hanmer 2009). Technology adoption and productivity in maize is thus closely tied to welfare.

The existing agricultural extension system in Malawi relies on Agricultural Extension Development Officers, henceforth extension agents, who are employed by the Ministry of Agriculture and Food Security (MoAFS). Many extension agents are responsible for upwards of 30-50 villages, which implies that direct contact with villagers is rare. According to the 2006/2007 Malawi National Agricultural and Livestock Census, only 18% of farmers participate in any type of extension activity. Extension agents cope with these staff shortages by relying on a small number of lead farmers, who are trained but not incentivized to disseminate knowledge via social learning.[18] Against this backdrop of staff shortages, maximizing the reach of social learning in the diffusion process may be a cost-effective way to improve the effectiveness of extension.

[17] For clarity, the model assumed that the potential net benefits of production were known to the farmer before deciding whether to become informed about the technology. In practice, farmers may or may not be aware that their private net benefits to adoption are high before becoming informed. Depending on the extent to which the farmer is *ex ante* aware that she has relatively high net benefits will determine whether this greater adoption is also associated with a greater propensity to become informed.

[18] The lead farmer model may additionally help farmers learn through allowing opportunity for frequent questions and conversations. Banerjee *et al* (2018) show that informing a subsample of individuals may lead to greater diffusion compared to broadcasting general information, because it creates opportunities for follow-up conversations.

### 3.2 Experimental Design

We first selected a new agricultural technology for each village (or "network") in our sample, and partnered with the Malawi Ministry of Agriculture to get their extension staff to train exactly two seed farmers in each village on the technology. Our experimental variation only changes how those seed farmers are chosen and holds all other aspects of the training constant. We identified the farmers in each of these villages who would be the "theoretically optimal" choices as seeds under specific formulations of the threshold model, where our objective is to maximize diffusion in the village over a 4-year horizon. Our four treatment arms randomly vary which theoretically optimal pair of seeds is trained in each village:[19]

1. Simple Contagion: Simple diffusion ($\lambda$=1) model applied to the network relationship data
2. Complex Contagion: Complex diffusion ($\lambda$=2) model applied to network relationship data
3. Geo Treatment: Complex diffusion ($\lambda$=2) model applied to network data constructed using only geographic proximity
4. Status Quo Benchmark: Extension worker selects the seed farmers based on her local knowledge

To implement this procedure, we first collected social network relationships data (to be described in detail in section 4) on the census of households in all study villages. The social network structures observed in these data allow us to construct network adjacency matrices for each of the 200 villages. Next we conduct technology diffusion simulations for all villages using these matrices, where each individual in the village draws an adoption threshold $\tau$ from the data, which is normally distributed[20] $N(\lambda, 0.5)$ but truncated to be strictly positive. We conduct simulations with $\lambda=1$ and $\lambda=2$ in all villages to evaluate simple and complex contagion respectively.

[19] In other words, we randomly assign "theories" or "threshold model formulations" to different villages. Randomization was stratified by district, and implemented using a re-randomization procedure which checked balance on the following covariates: percent of village using compost at baseline; percent village using fertilizer at baseline, and percent of village using pit planting at baseline. Randomization was implemented in each district separately.

In the simulations, when an individual is connected to at least $\tau$ individuals who are informed, he becomes informed in the next period. Once an individual is informed, we assume that all other household members are immediately also informed. We also assume that becoming informed is an absorbing state. As seed farmers are trained by extension agents, we assume all assigned seed farmers become informed.

We run the model for four periods.[21] Given the randomness built into the model, we simulate the model 2000 times for each potential pair of seeds in the village, and create a measure of the average information rate induced by each pair. We designate the pair that yields the highest average three-period information rate in our simulations as the two "*optimal seeds*" for each village for that particular model (simple contagion, $\lambda=1$ or complex contagion, $\lambda=2$). Armed with the identities of the optimal seeds under each model, we then randomly assign different villages in the sample to different models. The optimal seeds identified through the simple contagion ($\lambda=1$) simulation are trained on the technology in some randomly chosen villages assigned to treatment 1. Optimal seeds identified through the complex contagion ($\lambda=2$) simulation are instead trained in other villages that were randomly assigned to treatment 2.

To determine seeds for treatment arm 3, the simulation steps are the same as in the Complex Contagion case, except that we apply the procedure to a different adjacency matrix. To capture the idea that geography may be an easy way to capture key features of a social network, we generate an alternative adjacency matrix by making the assumption that two individuals are connected if their plots are located within 0.05 miles of each other in our geo-coded location data. We chose a radius of 0.05 miles because this characterization produces similar values for network degree measures in our villages as using the actual network connections measures.

[20] Heterogeneity in the model comes from variation across individuals in the net benefits realized by adopting pit planting. This affects the threshold number of connections an individual would need to have in order to get enough signals to be induced to adopt.

[21] We collected data for up to three agricultural seasons after the interventions were implemented, so our theoretical set-up matches our empirical research design. With knowledge of the value of $\lambda$, a policymaker could use the model to maximize adoption over any timeframe they cared about, either more short-term or more long-term.

The fourth group is the status-quo benchmark, where extension agents were asked to select two seed farmers as they normally would in settings outside the experiment. This benchmark constitutes a meaningful and challenging test for the simple and complex contagion treatments since the extension agents were able to use valuable information not available to researchers, such as the individual's motivation to take on the role. The benchmark treatment is similar to what the Malawi Ministry of Agriculture and other policymakers would normally do, so this is the most relevant counterfactual.[22]

Note that the Simple, Complex, and Geo seed farmer selection strategies were simulated in all 200 villages, so we know – for example – who the optimal simple contagion seed farmers would have been in a village randomly assigned to the complex contagion or the geo treatment. We label the counterfactual optimal farmers as "shadow seeds" or "shadow farmers".

## 4. Field Activities: Implementation of Interventions and Data Collection

### 4.1 Agricultural Technologies

In this section we describe the two technologies introduced to seed farmers and in section 4.1 we analyze data on crop yields to give further insights into the benefits of the technologies.

[22] Normally the Ministry only trains one "Lead farmer" per village, not two. In most villages, the Lead Farmer will already be established, except for villages in which there hasn't been an extension officer assigned to the village for a long time. The extension agents would have had to select a second seed farmer in benchmark villages due to the experiment.

*Pit Planting*

Maize farmers in Malawi traditionally plant seeds in either flat land or after preparing ridges. Ridging has been shown to deplete soil fertility and decrease agricultural productivity over time (Derpsch 2001, 2004). In contrast, pit planting involves planting seeds in a shallow pit in the ground, in order to retain greater moisture for the plant in an arid environment, while minimizing soil disturbance. The technique is practiced more widely in the Sahel, and has been shown to greatly enhance maize yields both in controlled trials and in field settings in East Africa, with estimated gains of 50-113% in yields (Haggblade and Tembo 2003, BenYishay and Mobarak 2014). In section 5.3 we offer further evidence on yield impacts in our sample of villages. The enhanced productivity is thought to derive from three mechanisms: (1) reduced tillage of topsoil, which allows nutrients to remain fixed in the soil rather than eroding, (2) concentration of water around the plants, which aids in plant growth during poor rainfall conditions, and (3) improved fertilizer retention.

Practicing pit planting may involve some additional costs. First, only a small portion of the surface is tilled with pit planting, and hand weeding or herbicide requirements may increase, though focus groups undertaken by the authors suggest that weeding demands were reduced substantially relative to ridging. Second, digging pits is a labor-intensive task with large up-front costs. However, land preparation becomes easier over time, since pits should be excavated in the same places each year, and estimates suggest that land preparation time falls by 50% within 5 years (Haggblade and Tembo 2003). BenYishay and Mobarak (2014) find that in Malawi, labor time decreases while the change in other input costs are negligible in comparison. Labor costs are minimized when pit planting is used on flat land.

*Crop Residue Management*

Seed farmers were also trained in crop residue management (CRM), a set of farming practices which largely focus on retention of crop residues in fields for use as mulch. Alternative practices commonly used by farmers include burning the crop residues in the fields and removing them for use as livestock feed and compost. The trainings emphasized the value of retaining crop residues as mulch to protect topsoil, reduce erosion, limit weed growth, and improve soil nutrient content and water retention. There is little experimental evidence on the impacts of CRM on soil fertility, water retention, and yields in similar settings.

### 4.2 Training of Seed Farmers

After we produced the lists of seed farmers for each village using the procedures described above, the extension agent assigned to the village trained the two seed farmers.[23] We provided extension agents with two seed farmer names for each village in experimental arms 1-3, and then replacement names if either of the first two refused to participate. Refusal was uncommon: we trained 93% of the selected seeds or their spouses. We conduct intent-to-treat analysis using the original seed assignment. The seed farmers received a small in-kind gift (valued at US$8) if they themselves adopted pit planting in the first year. There was no gift or incentive offered or provided on the basis of others' adoption in the village or the seeds' own adoption in subsequent years.

[23] As the technologies themselves were new, the extension agents were themselves trained by staff from the Ministry's Department of Land Conservation.

### 4.3 Data

After training the seed farmers, we collected up to three rounds of household survey data. Appendix Figure A1 shows the timeline of these data collection activities. We describe each major data source in turn.

*Social Network Census Data*

Targeting based on different network characteristics requires relatively complete information on network relationships within the village (Chandrasekhar and Lewis 2016). We reached more than 80% of households participating in the census in every sample village.[24]

The main focus of the social network census was to elicit the names of people each respondent consults when making agricultural decisions. General information on household composition, socioeconomic characteristics of the household, general agriculture information, and work group membership was also collected. Agricultural contacts were solicited in several ways: first by asking in general terms about farmers with whom they discuss agriculture. To probe more deeply, we also asked them to recall over the last five years if they had: (i) changed planting practices; (ii) tried a new variety of seed, for any crop; (iii) tried a new way of composting; (iv) changed the amount of fertilizer being used for any crop; (v) tried a new crop, such as paprika, tobacco, soya, cotton, or sugar cane; or (vi) started using any other new agricultural technology. If they responded affirmatively, we asked respondents to name individuals they knew had previously used the technique in the past and whether they had consulted these individuals. Finally we asked them if they discussed farming with any relatives, fellow church or mosque members, or farmers whose fields they pass by on a regular basis, or if there are any others with whom they jointly perform farming activities[25]. These responses were matched to the village listing to identify links. Individuals are considered linked if either party named each other (undirected graph), and all individuals within a household are considered linked.

[24] We interviewed at least one household member from 89.1% of households in Nkhotakota, 81.4% in Mwanza and 88.6% in Machinga. We interviewed both a man and a woman in about 30% of households.

[25] We also elicited their close friends and contacts with whom they share food, though we did not include these contacts as agricultural connections for the purposes of our network mapping.

*Sample Household Survey Data*

We collected survey data on farming techniques, input use, yields, assets, and other characteristics for a sample of approximately 5,600 households in the 200 sample villages. We attempted to survey all seed and shadow farmers in each village, as well as a random sample of 24 other individuals, for a total of about 30 households in each village.[26] In villages with fewer than 30 households, all households were surveyed. Three survey rounds were conducted in Machinga and Mwanza in 2011, 2012 and 2013, and two survey rounds were conducted in Nkhotakota in 2012 and 2013.[27] The first round asked about agricultural production in the preceding year—thus capturing some baseline characteristics—as well as current knowledge of the technologies, which could reflect the effects of training. Since the data was collected at the start of a given agricultural season, but after land preparation was complete, we observe three adoption decisions for pit planting for farmers in Mwanza and Machinga, and two decisions for farmers in Nkhotakota. Since crop residue management (CRM) decisions are made the end of an agricultural season after harvest, we observe CRM decisions for two agricultural seasons in Mwanza and Machinga, and one in Nkhotakota.

*Randomization and Balance*

Appendix Table A1 shows how observable characteristics from the social network census vary with the treatment status of the village. The table shows the results of a regression of the dependent variable listed in the column heading on indicators for the respondent residing in a benchmark, simple, complex, or geo treatment villages. District fixed effects are included in the regression, and standard errors clustered at the village level. P-values from tests comparing the different treatment groups as well as a joint test of all treatment groups are displayed. Few differences across treatment groups are statistically significant. Overall, the joint test reveals no differences for 10 out of 13 variables. Farm size, in column (9), is the most concerning: farmers in the benchmark villages have larger farm sizes on average than farmers in Simple and Complex villages, and the joint test across the treatment variables is significant at the 10% level. Additional analysis available from the authors controls for this variable in all specifications and finds that all results are robust to this control.

[26] In Simple, Complex and Geo villages there were 6 (2x3) seed and shadow farmers to interview, while in Benchmark villages there were 8 (2x4) seeds and shadows. Recall we do not observe Benchmark farmers in Simple, Complex and Geo villages.

[27] Unanticipated delays in project funding required us to start training of extension agents and seed farmers in Nkhotakota in 2012 instead of 2011 as we did in Mwanza and Machinga.

## 5. Empirical Results using Household-Level Data

Before reporting on the village-level experimental results, we establish some basic facts using household level data to help contextualize the results, and to show that the experiment was implemented as designed. We describe who the seeds are in each treatment arm using observable characteristics from the baseline household survey, and the rates at which the seeds adopt the new technologies themselves. We then show that the technologies we promoted on average improved agricultural yields. Next, we show that the seed farmers disseminate information on pit planting within the village. Finally, we show that individuals close to the trained seeds are more likely to adopt, so the individual-level adoption patterns are consistent with social learning.

### 5.1 Characteristics of the Seed Farmers under each Treatment

The simulations of the simple and complex contagion models generated different optimal seeds in most but not all cases. In 50% of villages, there was at least one seed who was judged as optimal in more than one (simple, complex or geo) model. Appendix table A2 describes the frequency of overlap in seeds across treatments. The most common scenario is that one simple seed is also a complex seed, which happens for about 25% of simple (and complex) seeds.[28] Optimal seeds are determined as a pair; in most cases this overlap occurs when there is one farmer who is both high degree and quite central. That farmer then becomes part of an optimal pair under simple contagion alongside a farmer who shares few connections with her in the network, and part of an optimal pair under complex contagion alongside a farmer who shares many connections with her. Even though the extension workers could have chosen central individuals, benchmark seeds are also simple seeds only 10% of the time and complex seeds only 12%. The least overlap is between the Geo seeds with all others. As expected, the simulations also generated different clustering patterns: 35% of our random household sample has a connection to a simple seed, and 6% are connected to both simple seeds. By contrast, 18% of households are connected to two complex seeds. For the geo-based seeds, 10% of households are connected to two seeds.

Table 1 describes differences in observable characteristics of seed farmers chosen under the four different targeting strategies. This table seeks to provide intuition in how the models differ in who is selected as a seed, so the analysis includes both actual seeds and potential (counterfactual) seeds (i.e. shadow farmers) to maximize sample size.[29] The most striking pattern in Table 1 is that the farmers selected as seeds under the geographic treatment are significantly poorer than other seeds. This is because many households live on one of their plots in Malawi. Households who are geographically close to lots of people will mechanically have less land, and these households tend to be poorer overall. Therefore while the idea of using geography as a proxy for one's network may be intuitive, the implications of geographic centrality may be context-specific, and inappropriate as a network-based targeting proxy in some cases.

[28] This is more common than what would be expected by chance. The median village in our sample has 58 households, so that 3.45% of households are seed farmers of each type. If all seed selections were random and independent from each other, then the probability that a seed of one type is also a seed of one of the three other types is $1 - (1 - .0345)^3 = .1$

[29] Table 1 is not demonstrating balance in the randomization of villages across treatment arms. Note that there are only 100 benchmark farmers since we never observe shadow benchmark farmers.

Seed farmers selected through the complex contagion simulations are the most "central" across all measures of network centrality we compute, including degree, between-ness and eigenvector centrality (columns 3-5).[30] Simple seeds have similar betweenness centrality as complex seeds, but lower eigenvector centrality.

Figure 1 shows examples of villages from our data with network links mapped and the locations of the simple, complex and geo seeds. They highlight the key difference between Simple and Complex targeting: in complex, the two seeds are either directly connected to each other, or have at least one common friend. In simple contagion, optimal seeds are spread out in order to reach more parts of the network quicker.[31] Geo seeds are generally close to one another, since the complex model was used in selecting the seeds, but are located in more peripheral locations within the network - as anticipated, given that they generally have less land and have low income, as shown in Table 1. Benchmark seed farmers are rarely very close to each other, such that they are unlikely to spark the diffusion process if decisions are governed by the complex contagion model.

### 5.2 Do Seed Farmers Adopt the Technology Themselves?

Panel A of Table 2 compares the technology adoption behavior of seed farmers to shadow farmers. We focus on this sub-sample, because shadow farmers act as the correct experimental counter-factual for the seed farmers to capture the causal effect of the intervention, removing any bias due to the seeds' position within their networks. We estimate the following equation, and Panel A displays the results:

$$y_{ivt} = \beta Seed_{ivt} + \delta_v + \epsilon_{ivt} \qquad (1)$$

where the dependent variable with an indicator for adoption, and $\delta_v$ are village fixed effects. Column (1) shows that trained seeds are 52% more likely in year 1 to know how to pit plant than shadow farmers. Shadow farmers' knowledge increases over the three agricultural seasons (from 16.5% in year 1 to 19% in year 2 to 29% in year 3) - as would be the case with technology diffusion within their villages - but seeds continue to have an informational advantage as seen in columns (2)-(3). Columns (4)-(6) show that seed farmers who are trained on pit planting adopt at a rate of 31-32% in all three years, compared to the low 5% adoption rate of shadow farmers in year 1.

---

30 Eigenvector Centrality is weighted sum of connections, where each connection's weight is determined by its own eigenvector centrality (like Google pagerank). Betweenness centrality captures that a person is important if one has to go through him to connect to other people. Therefore it is calculated as the fraction of shortest paths between individuals in the network that passes through that individual. See Jackson (2008) for more details.

31 The second simple farmer would be more central in a village which had multiple distinct cliques. However, we rarely observe this network structure in our data as almost all of our networks are organized around a giant component, similar to most empirical networks worldwide.

The trained farmers were also 14 percentage points more likely to try CRM in the first year after training (column 7). The rate for shadow farmers was high to begin with (32%), so managing crop residues was not as new or unfamiliar a concept as digging pits.[32] However, column (8) shows that CRM adoption declined quickly in the second year among both actual seeds and the shadows (from 46% to 26% for seeds), which is strong evidence that it was not deemed as useful a technology as pit planting. We therefore focus on pit planting in most of our empirical analysis, as the threshold model used to determine treatment does not allow for dis-adoption. We include CRM adoption results in Appendix tables A4 and A5.

[32] While pit planting is a new, largely unknown technology in Malawi (0.5% of farmers are practicing at baseline, and only 4.3% of farmers had heard of pit planting at the time of our census in the median village), farmers were using a range of strategies to deal with crop residues, including burning fields, leaving residues in fields, using residues as mulch, feeding residues to livestock, using residues to make compost, and, most commonly, burying residues in fields as they prepare new ridges. Several of these strategies overlap with the recommendations provided in our CRM training, creating measurement problems. Unlike pit planting, which is readily observable and distinct from other practices, whether farmers follow our CRM guidelines is not as easy to decipher in our data. Further, the optimal crop residue technique depends on household-specific factors like livestock owned, and farmers and extension experts disagree about the best practice. Ministry officials do agree that burning residue is a bad idea, and we observe burning frequency decreasing from 20% to 9% over the 3-year study period.

Panel B of Table 2 restricts the sample to only seed farmers (and drops all shadow farmers) and compares knowledge and adoption among seeds across the four experimental arms as follows:

$$y_{ivt} = \beta_0 + \beta_1 Simple_v + \beta_2 Complex_v + \beta_3 Geo_v + \delta X_v + \epsilon_{ivt} \quad (1)$$

Where $X_v$ include the re-randomization controls (listed in table notes), village size, the square of village size, and district fixed effects. Standard errors are clustered at the village level. Column (1) shows that in the first year, Benchmark seeds are most likely to say they know how to pit plant, while all other seeds are similar. The extension agents evidently chose seed farmers carefully to ensure that their chosen extension partners receive the initial training from them. However, in years 2 and 3, Simple and Complex seeds catch up and have similar levels of familiarity with pit planting as Benchmark seeds. Geo seeds continue to display lower familiarity in subsequent years.

Column 4 shows that there are no differences in adoption propensities across the four types of seeds in the first year. This implies that it is unlikely that any observed differences in village-wide adoption patterns across the four treatment arms that we will examine later, are driven by initial adoption differences inside the sub-sample of seed farmers. Columns (5) and (6) show that seed farmers in simple contagion villages become relatively more likely over time to adopt the technology. This could be due to the technology diffusion process, or in other words, a consequence of the experiment. Columns (7)-(8) show that there are no significant differences in adoption in seasons 1 or 2 for crop residue management.

### 5.3 Effect of Technology Adoption on Crop Yields

Table 3 compares yields between seed farmers to shadow farmers. We rely on the sub-sample of seeds and shadows to study yields, because seed farmers were the first to adopt the new technology, and Table 2 showed that there were large differences in adoption rates between seeds and shadow farmers. We estimate:

$$y_{ivt} = \beta Seed_{ivt} + \gamma X_v + \delta_t + \epsilon_{ivt} \quad (2)$$

where $y_{ivt}$ is log maize yields for farmer *i* in village *v* at time *t*, $Seed_{ivt}$ is an indicator for being the selected seed farmer, $X_v$ are control variables used during the re-randomization routine (see notes in Table 2), village size, village size squared, district fixed effects plus baseline land size. $\delta_t$ are year dummies. We use data from seasons 2 and 3. In the intent-to-treat specification in the first column, maize yields among seed farmers are 13% greater than the yields experienced by the shadow seeds. The fact that the technologies we promoted led to an increase in output strongly suggests that the information about pit planting that diffused through the networks was likely positive on average.

We report the local average treatment effect using an IV regression in the second column in which we instrument pit planting adoption with an indicator for being randomly assigned as the seed (rather than a shadow). In this specification, pit planting adoption is associated with a 44% increase in maize yield. However, we cannot rule out that CRM adoption also increased yields, potentially violating the exclusion restriction in the IV estimation.[33]

### 5.4 Seeds Farmers' Interactions with Other Villagers

In this section we examine whether seed farmers disseminate information about pit planting to their neighbors in the village. Table 4 uses data on conversations about pit planting that respondents had with others in the village. These conversations may arise from either the seed pushing information or from villagers eliciting information from the seeds. Each respondent was asked questions about seven other individuals in their village: whether they knew them, and what they had discussed. The seven individuals comprised of the two seed farmers, randomly selected shadow farmers, and a random sample of other village residents. We exploit the random variation from the experiment: for example, we compare the frequency of conversations with the complex seed farmers to the frequency of conversations with complex shadow farmers in other villages.[34]

[33] We also cannot rule out any labor or other input use response to training which may have positively contributed to yields without a profitability impact.

Table 4 shows that the experiment induced seed farmers to discuss pit planting with fellow villagers. Columns (1)-(3) show that there are more conversations with trained seeds than with shadows.[35] The simple contagion treatment led to more conversations with the simple partner, the complex contagion treatment led to significantly more conversations with the complex partner, and so forth.

Column (4) examines whether the treatments increased the total number of conversations about pit planting in the village. The dependent variable is equal to 1 if a respondent discussed pit planting with either seed farmers or one of the randomly selected individuals within the village. We find that respondents in Complex villages have a slightly higher likelihood of a having at least 1 conversation about pit planting compared to Benchmark or Geo villages.

### 5.6 Technology diffusion within the village

If adoption is a social contagion, individuals close to the seeds should be first to become informed and then adopt. To explore this, we estimate the following equation:

$$Y_{iv} = \alpha + \beta_1 1TSeeds_{iv} + \beta_2 2TSeeds_{iv} + \beta_3 1Simple_{iv} + \beta_4 2Simple_{iv} + \beta_5 1Complex_{iv} + \beta_6 2Complex_{iv} + \beta_7 1Geo_{iv} + \beta_8 2Geo_{iv} + \theta_v + \varepsilon_{iv}$$

Seeds and shadows are removed from the analysis. $1TSeeds$ is an indicator for the respondent being directly connected to exactly one seed farmer, and $2TSeeds$ indicates the respondent was directly connected to two seed farmers. Since network position is endogenous, we also control for whether an individual is connected to one or two Simple, Complex or Geo (actual or shadow) seeds, but these coefficients are not displayed in the table. Identification therefore comes from variation in closeness to the seed generated by the experiment. As an example, we can compare two farmers who are both connected to two 'Simple seeds', but where one farmer is in a village randomly assigned to the Simple treatment and his friend is trained, while the other was not.

[34] While all sample respondents in Simple treatment villages were asked about simple farmers, not all respondents in the remaining villages were, since we chose a random subset of shadow farmers. This is analogously true for complex and geo villages. We therefore flexibly control for the number of simple (complex, geo) farmers we asked about in the regression where the dependent variable is talking about pit planting with the simple (complex, geo) farmer.

[35] We may observe a treatment effect on conversations with the Simple partner in Complex villages and conversations with the Complex partner in Simple villages for one of two reasons: (i) as mentioned above, often there is one individual who would be chosen as a seed in both the simple and complex versions of the model but (ii) this may also be the outcome of the diffusion process. It is challenging to disentangle these two alternatives.

In the theoretical model, individuals have to become informed prior to adopting. As an empirical matter, it is unclear what level of knowledge is associated with "being informed" as used in the model. In table 5, we therefore consider three variables which represent increasing levels of information: whether the respondent has heard of pit planting; whether the respondent knows how to implement pit planting; and whether the respondent adopted pit planting (which implies not only knowledge but also that the signals that the respondent received were sufficiently positive). In season 1, the training led to more information transmission to those directly connected to seeds. In particular, those who have a direct connection to both seed farmers had the most knowledge. This is true for both measures of "knowledge": Whether the respondent had heard of pit planting and whether they reported being capable of implementing it. Respondents with two connections are 8.4 percentage points more likely to have heard of pit planting than those with no connection to a seed. This represents a 33% increase in knowledge relative to the mean familiarity among unconnected individuals. This effect is also statistically significantly different from the effect of being connected to one seed (p=.02). They are also 6.2 percentage points more likely to report knowing how to pit plant, a 108% increase over unconnected individuals and again significantly different from the effect of being connected to one seed (p = 0.072). These knowledge effects are suggestive of a complex contagion process ($\lambda = 2$) rather than simple contagion. The increased awareness of pit planting and knowledge of pit planting among households connected to two seeds persists into season 2 (columns 2 and 5), and two connections is again significantly more advantageous than one connection (p=.04 and .095, respectively).

We see no effect on adoption in the first year (column 7) among individuals directly connected to either one or two seeds. However, we do observe an adoption effect in year 2. This temporal pattern of results is consistent with the set-up of our theoretical model: Individuals become informed in year 1 and then some choose to adopt in year 2. Column (8) shows that households with two connections to trained seeds are 3.9 percentage points more likely to adopt in the second season than those with no connections, which represents a 90% increase in adoption propensity. Though the point estimate of the effect of 2 connections is considerably larger than the effect of a connection to one seed (3.9 pp compared to 1.2 pp), we cannot statistically reject that households with a connection to only one treated seed adopt less frequently (p=.17). We also observe that individuals who are within path length 2 of at least one seed (that is, a friend of a friend) are 2.2 percentage points more likely to adopt.

The predictions of the model for which individuals learn about pit planting are weakened as time passes and knowledge diffuses through the network. In all three of our dependent variables, this diffusion can be observed through large increases in knowledge and adoption over time in our omitted category: individuals with no direct connections to a seed. Among this group awareness increases from 22% to 39% from year one to three, while "knowing how" to pit plant increases from 6% to 15% and adoption increases from 1% to 4%. In principle, this diffusion should reduce power on our exogenous variation, as the number of connections to informed individuals becomes less correlated with the number of signals available to farmers. In practice, by year 3 we still see significance on the effects of two direct connections on one of our two knowledge variables ("knowing how" to pit plant, column 6), but we no longer see significant differences from direct connections in adoption or awareness of pit planting. Consistent with the hypothesis that this loss in precision is due to diffusion in the network, we see that adoption increases among those at moderate distance to the seeds in year 3: column (9) shows that households within path length 2 are more likely (3.7 pp) to have adopted over those who are socially more distant.[36]

In summary, analysis using individual-level data demonstrates that individuals who are initially close to the trained seeds are more likely to adopt than individuals with no direct connections – as one would expect if the experiment is inducing social network-based diffusion. The data also suggest that having two direct connections – and not just one – is important for diffusion. This is suggestive evidence in favor of the complex contagion model: farmers may need to know multiple informed connections before becoming informed, and then subsequently adopting, themselves.

## 6. Village-Level Experimental Results: Does Theory-based Targeting increase Adoption?

In this section, we report experimental results on village-level outcomes across the four types of study villages, and use them to test the predictions of simple and complex contagion theory. We measure technology adoption in our surveys, because this is ultimately of most interest to policy, and because adoption can be observed and measured most precisely than being informed.

### 6.1 The Advent of Diffusion under Simple and Complex Contagion

To generate specific theoretical predictions, we first assume that simple contagion ($\lambda=1$) is the right model, and under this assumption, we simulate an indicator for "informed about new technology" for every sample household in every one of our 200 villages for three seasons after the experiment was implemented. This allows us to create a simulation of the adoption patterns that we should observe in *every* village, if the simple contagion model correctly describes technology diffusion in our setting. Next, we repeat the same exercise, but under the assumption that complex contagion ($\lambda=2$) is the right model, to generate predictions for the technology diffusion we should observe if instead complex contagion theory correctly describes the diffusion process. We then compare the actual adoption data to these simulated predictions.

[36] This is a lower power test of the model than the direct connections test as it is imperfectly correlated with the number of informed, indirect connections to seeds (which is unobserved). We do not see a significant effect of this variable on knowledge outcomes, though coefficients are positive.

One key feature of the threshold model that helps distinguish complex from simple contagion is that for almost *any* choice of seed farmer, the diffusion process will start under simple contagion. However, if the diffusion process is complex, then many potential pairs of seeds would never generate any social diffusion. This is because when two seeds are not proximate to each other in the network map and they don't share any common connection, then no other individual is connected to multiple informed seeds, and the technology never diffuses. This leads us to focus on the *advent* of diffusion in our sample villages as a key outcome. We define "any adoption" as an indicator for villages which have at least one household (other than the seeds) that adopted pit planting. Our models actually simulate being "informed" and not adoption directly, but in order to be parsimonious and tractable we compare the rates of being informed from the simulations to adoption rates in the data.[37] The focus on "any adoption" yields a sharp prediction to distinguish complex contagion from the other treatments: If complex contagion is the correct description of the diffusion process, then the indicator "any adoption" should be significantly higher under the complex treatment than all other treatments. Note that the analogous theoretical prediction does not exist for the simple contagion model, because all four treatments are likely to see some advent of diffusion if the world is simple.

The left part of Figure 2 shows the *predicted* fraction of villages with "any adoption" from simulating the model for all sample villages when $\lambda=1$ (Simple contagion) and $\lambda=2$ (Complex contagion).[38] Since the goal is to compare these simulations to the actual data, we design the simulations to reflect the fact that we only observe a random sample of households in these villages.[39] The right part of Figure 2 shows the empirical counterpart: "any adoption" rates in the data in years 2 and 3.

[37] Simulating adoption in the model would require a number of additional assumptions, including estimates of signal accuracy, the distribution of net benefits, and any heterogeneity in prior beliefs which may exist. Being informed is necessary but not sufficient to adopt.

[38] These simulations exclude 12 villages where at least one of the extension worker chosen seeds (benchmark) was not observed in our social network census. This occurred because the spatial boundaries of villages are not always clearly delineated in Nkhotakota.

When the threshold is set to λ=1, diffusion is predicted to be widespread. In year 2, 85% of villages where Geo and Benchmark partners were trained are predicted to have some sampled diffusion, and that rate goes up to 94% with Simple and Complex partners. The predicted rates of 'any diffusion' are even higher in year 3.

The risk of no diffusion increases if the diffusion process is characterized by complex contagion. In that case, the model predicts that more than half of the villages assigned Simple, Geo or Benchmark partners will not see any sampled diffusion at all in year 2. In contrast, when complex seeds are trained, 70% of villages are predicted to experience some diffusion.

Comparing the theoretical simulations to the data on the right side of Figure 2 shows that the data are more consistent with the patterns generated by a complex (rather than simple) learning environment in three distinct ways. First, the simple contagion simulations suggest that we should observe a much higher fraction of villages with some adoption than is true in the data. Second, simple contagion predicts that the "any adoption" outcome should not be very sensitive to the identity of the seed farmer who is initially trained. In contrast, the identity of the seed farmer dramatically alters this outcome in the data. Finally, the complex contagion simulations predict that the complex partners will maximize the fraction of villages with some adoption, which is exactly what we observe in the data.

The first two columns of Table 6 replicate the data panels on the right side of figure 2 in a regression framework. The propensity for "Any Adoption" in season 2 in statistically significantly larger in villages assigned to the complex contagion treatment relative to Benchmark villages. The 25 percentage point gap is large relative to the "any adoption" rate of 42% in our Benchmark villages. The 'any adoption' rate in complex villages is also 15 percentage points larger than in Geo villages (p-value = 0.10) and 10 percentage point larger compared to villages assigned to the simple contagion treatment (p-value = 0.30). In season 3, Simple, Complex and Geo villages all attain a statistically larger rate of "any adoption" than Benchmark villages. 85% of Complex villages had at least one non-seed adopter, compared to 73% of Simple and Geo villages and 54% of Benchmark villages.

[39] The simulations use the full social network to predict becoming informed, measured here through adoption. We then sample from the full network to better mimic our data. In the model, the rate of any adoption is identical in years 2 and years 3. If there was no adoption by year 2, there is no way there will be any additional adoption taking place in year 3. The sampling process, however, generates the increase over time observed in the figure. If the rate of adoption is low, as is empirically the case, then a random sample may miss all adopters. As the number of adopters increases over time, the random sample is more likely to pick up an adopter and hence the rate of any adoption increases over time in the figure.

### 6.2 Adoption Rates across Treatment Arms

Columns 3 and 4 in Table 6 document treatment effects on the adoption rate, which is defined as the proportion of farmers who adopted pit planting in each agricultural season. Both simple and complex contagion villages have higher adoption rates relative to the benchmark in season 2. Compared to the benchmark rate of 3.8%, complex and simple villages both experience a 3.6 percentage point higher adoption rate. We cannot reject that the adoption rates are the same in Simple, Complex and Geo villages. The adoption rate increases across all four types of villages in season 3. The adoption rate increases in the benchmark villages, the reference category, from 3.8% to 7.5% from season 2 to 3. With the smaller sample size of 141 villages in season 3, we cannot reject that the adoption rate is the same across all treatment types, though the point estimate on Complex remains the largest, and is equal in magnitude to the effect size observed in season 2.

Appendix Table A3 shows the results of analogous regressions on "data" generated from the theoretical simulations we conducted to create the left panels of Figure 2.[40] The simulations predict that the complex treatment should perform best both in terms of "any adoption" and the "adoption rate" if the learning environment in reality is complex. If the learning environment is instead simple, then we should expect to see few statistical differences in diffusion across targeting strategies by season 3, since the choice of seed partners is relatively unimportant if the technology diffuses easily. These patterns are broadly consistent with what we observe in the data: the diffusion process is far too slow to be consistent with simple contagion. Our parameterization of $\lambda$=2 does not provide a perfect fit for the data. For example, the simulations in Appendix Table A3, columns 2 and 4 suggest that the complex treatment should produce a larger adoption rate than the simple treatment if the learning environment is complex. In Table 6, we cannot statistically distinguish between these two treatments. Overall, however, the empirical results in Tables 6 and in Figure 2 appear more consistent with a complex learning environment than with simple contagion.

[40] We theoretically simulate the rate of "becoming informed" about the technology, and this is not as good a proxy for the "adoption rate" as the simulated "at least one person becoming informed" is for the "any adoption" variable. We therefore need to be cautious about comparing columns 3 and 4 across Table 6 (the data) and Appendix Table A3 (the simulations).

### 6.3 Heterogeneity in the Learning Environment

Our theoretical micro-foundation suggests that the threshold model describes diffusion as a learning process where farmers need to aggregate signals and ultimately adopt if those signals are sufficiently positive. Thus, we anticipate that our treatments will be most effective in inspiring adoption for farmers who are likely to receive positive signals.[41] We use two different approaches to identify subsets of sample farmers for whom the signal about the technology's profitability is more likely to have been "good", and we use such farmers to construct empirical tests. First, the Ministry of Agriculture recommends pit planting only for flat land, and labor costs of pit planting are lower on flat land.[42] Focus group discussions in our sample villages confirmed that villagers thought pit planting was more suitable for flat rather than sloped land. We therefore expect more positive treatment effects for farmers who possess land that is flat and not sloped.

---

[41] This prediction distinguishes the threshold model from diffusion models based on imitation or infection, but not necessarily from other diffusion models that are also based on learning.
[42] Pit planting is possible on land with some slope, but in those cases, the pits need to be constructed differently, and our extension workers were not trained on that technique.

Second, pit planting is in general a new technology in Malawi, but there is heterogeneity across villages in how novel it is. The information environment should be most affected by our treatments when the technology is truly novel, both because each piece of new information will have a larger effect on posterior beliefs, and because the differences between treatments may become less stark if some farmers in the network are already informed about pit planting.[43] For these reasons, we anticipate larger treatment effects in villages where the technology is truly novel.

Table 7 explores the heterogeneity in treatment effects across these two dimensions, by interacting the randomized treatments with an indicator for "Farmer likely to receive a Good Signal". This "Good Signal" variable is first defined as the farmer having flat land in columns 1 and 2, and then re-defined as "Village with lower-than-median familiarity with the technology at baseline" in columns 3 and 4. "Bad signal" refers to the converse of these characteristics. The equation estimated:

$$y_{ivt} = \beta_0 + \beta_1 Simple_v * Bad\ Signal + \beta_2 Complex_v * Bad\ Signal + \beta_3 Geo_v * Bad\ Signal + \beta_4 Good\ Signal + \beta_5 Simple_v * Good\ Signal + \beta_6 Complex_v * Good\ Signal + \beta_7 Geo_v * Good\ Signal + \delta X_v + \epsilon_{ivt}$$

The reference group comprises of farmers who are likely to receive a bad signal in Benchmark villages. Our hypothesis is that among those who receive a positive signal, we will observe more diffusion in Complex villages if the true model is Complex.

Columns (1) and (2) show that adoption in year 2 is higher for farmers who have flat land in Simple, Complex and Geo villages compared to farmers with flat land in Benchmark villages. In year 3, we see that Complex villages continue to have a larger adoption rate than Benchmark villages for farmers with flat land.

Columns (3) and (4) show that the complex treatment performs best in villages where the technology was relatively novel. In this sub-sample, the adoption rate is statistically significantly higher in Complex Contagion treatment villages compared to both the Simple contagion and the benchmark treatments in year 3.

[43] Given that there is approximately 0% adoption at Baseline, it is additionally unlikely that previously informed farmers are disseminating a positive signal about this technology.

To summarize, these heterogeneity tests indicate that targeting based on complex contagion is most effective precisely in the types of villages and for the types of farmers where we theoretically expect it to. Just as importantly, those same results are not replicated for the subsets of villages and farmers where the model of social learning is not as good a fit. We interpret these tests as strongly suggesting that the social learning environment about agriculture in rural Malawi is well characterized by complex contagion. The policy implications that stem from that observation are (a) The network position of who you initially target with the new technology matters, and (b) Complex contagion-theory-based targeting can improve the speed and scope of technology diffusion relative to other forms of targeting.

## 7 Cost-effective, Policy-Relevant Alternatives to Data-Intensive Targeting Methods

While targeting based on the threshold model improves technology diffusion, eliciting the social network map to achieve these gains is expensive. Our geography-based treatment arm was an attempt to assess how much of the diffusion benefit derived from applying network theory could be achieved without having to resort to expensive data collection methods (since each household's physical location is much easier to observe than network relationships). That specific approach was not an unqualified success. Table 1 provides some insights as to why. Even though two Geo seeds are often clustered together, in this setting the seeds are poorer, and have fewer connections to others in the network.

Combining our experimental results with research on other inexpensive procedures to identify the optimal seeds under complex contagion theory would make network-based targeting more policy relevant and scalable. In some contexts, relevant groups within the village may be well-known by policy-makers, and our result would suggest that the critical goal for policy is to saturate individual groups with a few trained seeds. For other contexts, we may need to infer more about the network, and a few recent papers have suggested promising, less expensive methods for inferring network characteristics. Banerjee et al (2018) suggests that despite the implicit challenges in learning about network structure, the simple question of "if we want to spread information about a new loan product to everyone in your village, to whom do you suggest we speak?" is successful in identifying individuals with high eigenvector centrality and diffusion centrality, who ultimately improve the diffusion process. Breza et al (2017) suggest that aggregate relational data collected from a smaller sample combined with a census can yield accurate estimates of network characteristics.

While we cannot test the viability of either approach empirically, we can explore via simulations some alternate strategies that extension officers could use to identify useful partners. We make use of the fact that – unlike other network statistics – "degree" of a network node (i.e. simply the number of direct connections it has to other nodes) can be estimated from a single interview (Chandrasekhar and Lewis 2016). In this section, we simulate the effects of several potentially low-cost strategies in our data, assuming a complex contagion learning environment.

We suppose that an extension agent enters a village and randomly selects a small number of farmers to interview, and only asks one question from our social network census: "Do you discuss agriculture frequently with anyone in the village? What is the name of the person you speak with about agriculture frequently?" The response to this question generates a small list of names. The extension agent can then use the responses to the initial interviews to select any follow-up interviews.

We simulate six candidate targeting strategies, discussed in the appendix. While most do not perform competitively with the optimal complex contagion targeting, we find that strategies which leverage the highest degree respondent from the random sample can approach the performance of the optimal targeting. More specifically, if we train any two connections of that respondent, we achieve 73% of the optimal adoption rate with just 2 total interviews. If the extension agent then identifies her two highest degree friends (which requires an additional 5 or so interviews to determine which connections are the highest degree), and trains those two, we simulated that those trained seeds would achieve 84-90% of the adoption under optimal complex partners depending on the number of initial interviews. The intuition is that in a complex learning environment, it is most useful to identify two seeds who have at least one connection in common, and who are located near the center of the largest inter-connected component of the network. Training two high degree friends of the highest degree farmer guarantees at least one high degree person will become informed, and generates a high likelihood of creating other connections in common.

These simulations therefore suggest that it is possible for us to learn about the relevant pieces of network structure to enhance technology diffusion under a complex contagion learning environment at modest cost.

## 8 Concluding Remarks

This paper sought to use a theory of social learning to increase diffusion of a new technology. We first develop a theory-driven methodology to select seed farmers who are predicted to maximize diffusion of information about a productive new agricultural technique in Malawi under different theoretical assumptions about the process of diffusion. We then implement those choices using field experiments on agricultural extension conducted in partnership with the Malawi Ministry of Agriculture. This allowed us to test whether (a) Theory-driven targeting using detailed social network data can increase technology adoption relative to the status quo approach to agricultural extension services; (b) A less data-intensive approach can generate similar gains; (c) The learning environment for new technologies is "complex" or "simple", in the language of the linear threshold model; and (d) whether the diffusion we measure follows a learning process such as the one suggested by the threshold theory.

The theory and simulations provide a few specific predictions for the experiments and data. First, targeting is only critical if the learning environment is complex since under simple learning all treatments will generate similarly large adoption gains after three years. Second, if diffusion has complex contagion properties, then it is useful to cluster seeds in the same part of the network; otherwise no one crosses the threshold and we would observe no adoption. Third, under complex learning, multiple connections to seeds should be predictive of adoption. Our estimates suggest that most farmers are convinced to adopt a new technology only if they receive information about it from multiple sources. This implies that diffusion follows a Complex Contagion pattern. Our simulations suggest that with only about 10 interviews per village, it may be possible to identify individuals who can trigger the diffusion process.

Designing experimental treatments tightly linked to specific pre-formulated theoretical structures conveys several advantages. First, the use of the model *ex ante* in designing treatments commits us to testing a particular model. This eliminates the possibility of searching over potential theoretical models to *ex post* rationalize surprising (and possibly spurious) patterns in the data.

Second, as our treatment arms themselves incorporate the structure of the theoretical model, we can use these "structural experiments" to directly demonstrate the implications of simple or complex contagion models. We present simulated counterfactuals alongside our actual empirical results to allow readers to view what the experiment revealed about simple or complex contagion.

Third, while our approach is stylized, it may allow for a greater degree of external validity, because the theories we test were formulated independent of context. The results of other attempts in the literature to diffuse new technologies through social networks[44] may be context-dependent because they rely on local institutions – such as local leaders or focus groups – to identify network partners. While learning may not follow a complex contagion pattern in other settings, the use of theory led to treatments that are transparent in the objective used to select individuals.

[44] Kremer et al (2011) identify and recruit 'ambassadors' to promote water chlorination in rural Kenya, Miller and Mobarak (2014) first markets improved cookstoves to 'opinion leaders' in Bangladeshi villages before marketing to others, and BenYishay and Mobarak (2015) incentivize 'lead farmers' and 'peer farmers' to partner with agricultural extension officers in Malawi.

## Appendix

*A.1. Simulations of adoption regressions in section 6.2*

Appendix table A3 reflects the adoption patterns we should observe under the parameterizations of the simple and complex learning environments we used in identifying our treatment partners. Table A3 presents these simulation results for two different measures of technology adoption: the adoption rate and an indicator for villages with any non-seed adopters. We predict these outcomes for all four experimental arms that were implemented in the field. Panel A shows what we should expect to observe across treatments based on simulations of the model with $\lambda=1$ (Simple contagion), and Panel B reports predictions under $\lambda=2$ (Complex contagion).[45] In all cases we use the simulated contagion (becoming informed) to predict adoption outcomes as described in the main text.

Columns (1)-(2) show the results for adoption rate outcomes. Complex partners initially maximize adoption in year 2 even if the learning environment is simple, but in year 3 adoption rate is highest when the simple seeds are trained. However, the effects of training simple and complex seeds are not statistically distinguishable (p=.73) for these outcomes simulated under simple contagion. Under simple contagion, villages where the Geo seeds are trained exhibit the lowest adoption rates. Columns (3)-(4) show a very similar set of results for whether anyone adopts under simple contagion. Taken together, these results indicate that the simple treatment is not expected to dominate alternative targeting strategies even if the contagion process is simple. This reinforces the intuition that if farmers truly have a low threshold for adoption, the diffusion process is not likely to be particularly sensitive to who is initially targeted with information.

[45] The table differs from Figure 1 in two key dimensions: (1) this uses the realized randomization and not all villages as in figure 1, and (2) includes additional stratification control variables as in the empirical analysis.

In contrast, when we conduct simulations assuming the complex contagion model is correct, the complex treatment is predicted to increase adoption significantly more than all other treatments (Panel B of Table A3). The Complex treatment out-performs the simple, Geo and Benchmark treatments in terms of all adoption outcomes during both years (with statistical tests for differential effects producing p-values below 0.001 for every comparison).

*A.2. Simulation of cost-effective targeting strategies*

For our simulations, we suppose that our extension agent starts with a random sample of candidate respondents, and is able to screen out individuals with less than 2 connections.[46] We suppose the extension agent starts with a list of 2-10 total random farmers.

Starting from that random sample of farmers, we solicit each farmer's connections and calculate each random farmer's degree. We then focus on 6 candidate targeting strategies:

A. Trains two randomly selected people from that list (used as a benchmark)
B. Trains the two highest degree people from that list
C. Select two random friends of the highest degree person from that list
D. Trains the two highest degree connections of the highest degree farmer from the random sample (requires interviewing all connections of the highest degree respondent to determine their degree)
E. Selects two farmers from that list at random; interviews two of their connections (selected at random) and trains two of connections' connections[47]
F. Trains the highest degree respondent and one of his connections (at random).

[46] Since these individuals are quite socially isolated, we do not anticipate that such a screening would be particularly challenging; if not it may add modestly to the implementation costs as the agent must pose a screening question to identify the subset of connected individuals from their random sample.

[47] This "friends of friends" approach to identifying central people was inspired by Kim et al (2015), Feld (1991) and Christakis and Fowler (2010), who note that randomly selected connections tend to be more central than randomly selected nodes in a network. We again assume that the extension agent is able to screen out potential trainees with less than two total connections.

For each of these five candidate strategies, we simulate adoption rates after 4 rounds of simulations against the seeds chosen by our complex contagion treatment. We find that Strategy A, selecting two farmers at random, achieves 57% of the adoption produced by the complex contagion treatment. We can then view the other targeting strategies in terms of their performance above the random benchmark. Strategy B is identical to random selection with only 2 initial interviews, and so similarly generates 57% adoption; however, as the extension agent interviews more people to identify these high degree individuals it performs somewhat better, achieving 70% of the complex contagion adoption with 10 total interviews. Strategies C and D both leverage the highest degree respondent from the initial random sample. These perform the best out of the strategies we consider. Strategy C achieves 73% of the optimized adoption with just two total interviews, which increases modestly to 76% of the optimized adoption as the number of interviews grows to 10 to better identify a high degree individual. Strategy D, our best performing strategy, achieves 84% of the optimized adoption with 2 initial interviews (necessitating 8 total interviews as the connections are interviewed), and up to 90% of the optimized adoption with 8 initial interviews (and 13 total interviews). Strategy E requires a total of 4 interviews, and achieves 69% of the optimized adoption. Strategy F achieves 60% of the optimized adoption with 2 interviews, and up to 67% of optimized adoption with up to 10 interviews.

Clearly the most effective strategies are those that identify a high degree farmer and train her connections. Given the nature of the complex contagion learning process, the intuition is clear: training two high degree friends of someone who is high degree means that three people with many connections in the same part of the network will become informed. With clustered networks, it is likely that others will as well.

Figure 1: Network Maps

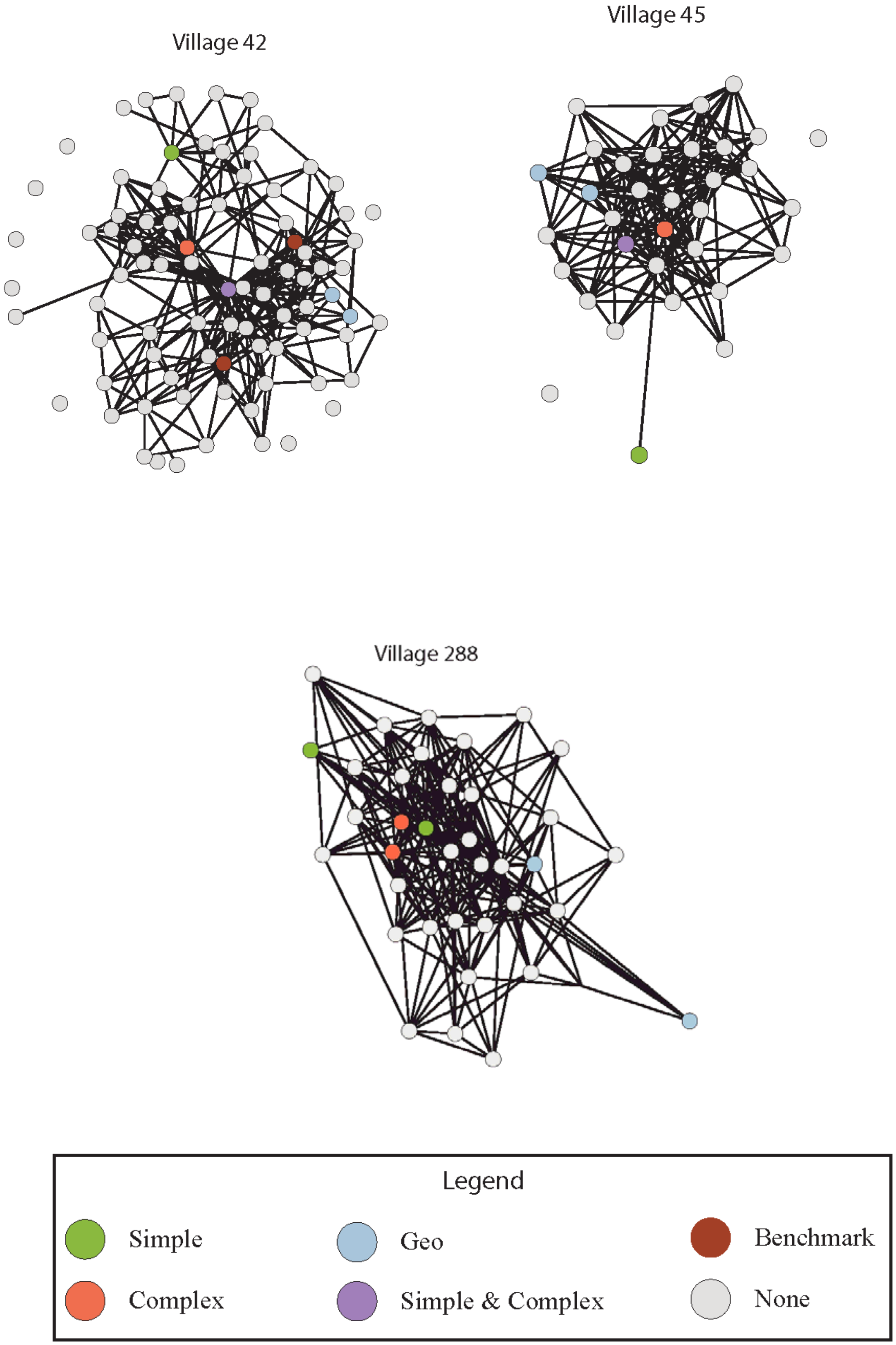

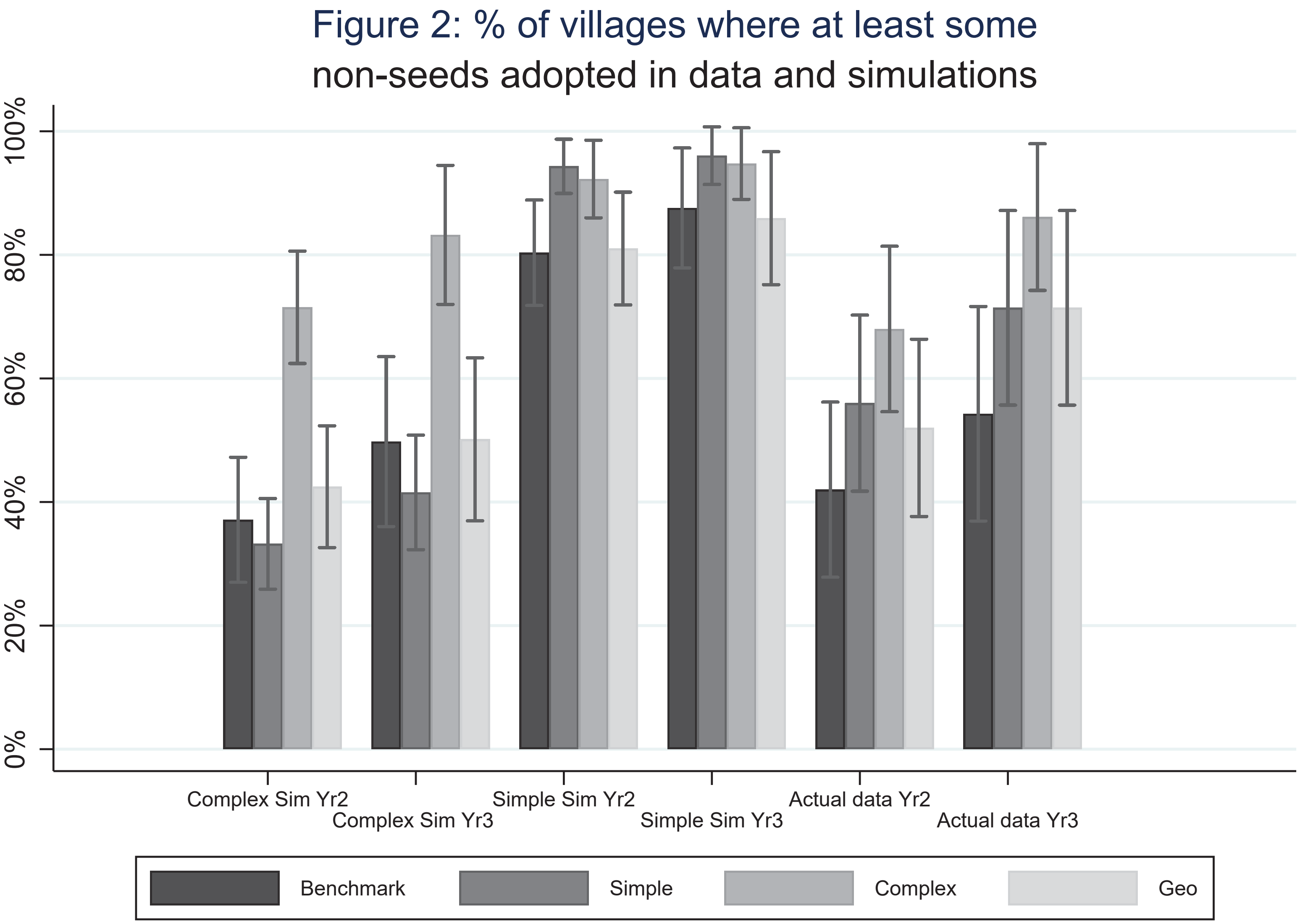
Figure 2: % of villages where at least some
non-seeds adopted in data and simulations
100%
80%
60%
40%
20%
0%
Complex Sim Yr2
Complex Sim Yr3
Simple Sim Yr2
Simple Sim Yr3
Actual data Yr2
Actual data Yr3
Benchmark
Simple
Complex
Geo

Table 1: Characteristics of the Seeds Chosen by Each Treatment Arm

| | Wealth Measures | | | | Social Network Measures | | | | | |
|---|---|---|---|---|---|---|---|---|---|---|
| | Farm Size | | Total Index (PCA) | | Degree | | Betweenness Centrality | | Eigenvector Centrality | |
| | (1) | | (2) | | (3) | | (4) | | (5) | |
| Treatment arm: | | | | | | | | | | |
| Simple Contagion | -0.152 | | 0.113 | | 0.455 | | 156.009 | ** | 0.009 | |
| | (0.19) | | (0.23) | | (1.03) | | (67.93) | | (0.01) | |
| Complex Contagion | -0.037 | | 0.380 | | 3.725 | *** | 146.733 | ** | 0.064 | *** |
| | (0.19) | | (0.23) | | (1.02) | | (67.74) | | (0.01) | |
| Geographic | -0.614 | *** | -0.740 | *** | -3.616 | *** | -90.204 | | -0.046 | *** |
| | (0.19) | | (0.23) | | (1.03) | | (68.04) | | (0.01) | |
| **p-values for tests of equality in seed characteristics** | | | | | | | | | | |
| Simple = Complex | 0.335 | | 0.067 | | 0.000 | | 0.815 | | 0.000 | |
| Complex = Geographic | 0.000 | | 0.000 | | 0.000 | | 0.000 | | 0.000 | |
| Simple = Complex = Geographic | 0.000 | | 0.000 | | 0.000 | | 0.000 | | 0.000 | |
| N | 1248 | | 1248 | | 1232 | | 1232 | | 1232 | |
| Mean Value for Seeds in Benchmark Treatment (omitted category) | 2.06 | | 0.626 | | 11.9 | | 169 | | 0.173 | |
| SD for Seeds in Benchmark Treatment | 2.97 | | 1.7 | | 6.77 | | 343 | | 0.0961 | |

Notes

1 The sample includes all seeds and shadows. The sample frame includes 100 Benchmark farmers (2 partners in 50 villages), as we only observe Benchmark farmers in Benchmark treatment villages, and up to 6 additional partner farmers (2 Simple partners, 2 Complex partners, and 2 Geo partners) in all 200 villages.

2 Benchmark treatment seeds are the omitted category.

3 *** p<0.01, ** p<0.05, * p<0.1

Table 2: Seed knowledge and adoption

| | Knows How to Pit Plant | | | Adopts Pit Planting | | | Adopts CRM | |
|---|---|---|---|---|---|---|---|---|
| | (1) | (2) | (3) | (4) | (5) | (6) | (7) | (8) |
| **Panel A** | | | | | | | | |
| Seeds | 0.518 *** | 0.367 *** | 0.245 *** | 0.258 *** | 0.230 *** | 0.182 *** | 0.137 *** | 0.047 |
| | (0.04) | (0.04) | (0.05) | (0.03) | (0.03) | (0.04) | (0.04) | (0.04) |
| Seasons | 1 | 2 | 3 | 1 | 2 | 3 | 1 | 2 |
| N | 659 | 735 | 503 | 686 | 672 | 489 | 686 | 467 |
| Mean of Shadows | 0.165 | 0.187 | 0.291 | 0.0541 | 0.0929 | 0.139 | 0.32 | 0.207 |
| SD of Shadows | 0.371 | 0.39 | 0.455 | 0.227 | 0.291 | 0.347 | 0.467 | 0.406 |
| **Panel B** | | | | | | | | |
| Simple Contagion | -0.133 * | -0.067 | 0.108 | -0.006 | 0.129 * | 0.176 ** | 0.078 | -0.097 |
| | (0.07) | (0.07) | (0.08) | (0.07) | (0.07) | (0.09) | (0.08) | (0.09) |
| Complex Contagion | -0.120 * | -0.058 | 0.007 | -0.020 | 0.002 | 0.037 | -0.001 | -0.077 |
| | (0.07) | (0.07) | (0.08) | (0.08) | (0.07) | (0.08) | (0.08) | (0.09) |
| Geographic | -0.193 *** | -0.255 *** | -0.150 | -0.095 | -0.064 | -0.003 | -0.011 | -0.075 |
| | (0.07) | (0.07) | (0.09) | (0.08) | (0.07) | (0.08) | (0.08) | (0.10) |
| Seasons | 1 | 2 | 3 | 1 | 2 | 3 | 1 | 2 |
| N | 343 | 383 | 264 | 353 | 352 | 259 | 353 | 243 |
| Mean of Benchmark | 0.824 | 0.653 | 0.547 | 0.337 | 0.276 | 0.238 | 0.442 | 0.339 |
| SD of Benchmark | 0.383 | 0.479 | 0.502 | 0.476 | 0.45 | 0.429 | 0.5 | 0.478 |
| *p-value for tests of equality in adoption rates across treatment cells:* | | | | | | | | |
| Simple = Complex | 0.872 | 0.904 | 0.242 | 0.862 | 0.0766 | 0.108 | 0.311 | 0.808 |
| Complex = Geographic | 0.377 | 0.0155 | 0.111 | 0.36 | 0.358 | 0.625 | 0.886 | 0.977 |
| Joint test of 3 treatments | 0.472 | 0.0206 | 0.0109 | 0.252 | 0.00846 | 0.049 | 0.235 | 0.795 |

Notes

1 In Panel A, all columns compare seed farmers to shadow farmers. Village fixed effects are included, and standard errors are clustered at the village level.

2 In Panel B, the sample includes only seed farmers, and the reference group is Benchmark seed farmers. The specification also includes controls which were used in the re-randomization routine (percent of village using compost at baseline; percent village using fertilizer at baseline; percent of village using pit planting at baseline); village size and its square; and district fixed effects. Standard errors are clustered at the village level.

3 *** p<0.01, ** p<0.05, * p<0.1

Table 3: Yields of Actual Seeds Relative to Shadow (Counterfactual) Farmers

| | Log of Agricultural Yields | | | |
|---|---|---|---|---|
| | (1) | | (2) | |
| Seed | 0.126 | ** | | |
| | (0.061) | | | |
| Adopted PP | | | 0.443 | ** |
| | | | (0.210) | |
| N | 959 | | 959 | |
| Mean of Shadows | | | | |
| Season | 2,3 | | 2,3 | |

Notes

1 The specification also includes district and season fixed effects and controls for total farm size, village size, and village baseline usage of fertilizer, composting and pit planting. Agricultural yields were winsorized.

2 Sample includes only seed and shadow (counterfactual) farmers. Benchmark villages are excluded. Standard errors are clustered at the village level.

3 *** p<0.01, ** p<0.05, * p<0.1

Table 4: Conversations other farmers report having about Pit Planting with Seed and Shadow Partners

| | with Simple Partner | | with Complex Partner | | with Geo Partner | | At least 1 conversation with seeds or randomly chosen villagers | |
|---|---|---|---|---|---|---|---|---|
| Treatment arm: | (1) | | (2) | | (3) | | (4) | |
| Simple Contagion | 0.085 | *** | 0.043 | ** | 0.009 | | 0.034 | |
| | (0.026) | | (0.019) | | (0.016) | | (0.026) | |
| Complex Contagion | 0.055 | *** | 0.097 | *** | 0.011 | | 0.052 | * |
| | (0.020) | | (0.024) | | (0.016) | | (0.026) | |
| Geographic | -0.003 | | 0.008 | | 0.050 | ** | -0.021 | |
| | (0.021) | | (0.020) | | (0.020) | | (0.026) | |
| N | 10354 | | 10712 | | 10585 | | 11606 | |
| Mean of Benchmark | 0.176 | | 0.189 | | 0.152 | | 0.370 | |
| SD of Benchmark | 0.381 | | 0.391 | | 0.359 | | 0.483 | |
| *p-values for equality in coefficients:* | | | | | | | | |
| Simple = Complex | 0.209 | | 0.023 | | 0.9 | | 0.414 | |
| Complex = Geo | 0.003 | | 0 | | 0.04 | | 0.001 | |
| Simple = Geo | 0 | | 0.083 | | 0.044 | | 0.016 | |
| Season | All | | All | | All | | All | |

Notes

1 Sample excludes seed and shadow farmers.

2 Also included are controls as listed in table 2 note 2 and district fixed effects. In columns (1)-(3), we also include controls for the number of partner farmers (of the type asked about in the respective column) we asked about in the questionnaire by including a dummy variable for each number of partner farmers from 0 to 4. In column (4) we also include controls for the number of seeds we asked respondents about and the number of randomly selected villagers. This varies by village treatment type, since we do not know observe shadow benchmark villages in non-Benchmark villages, and in those villages were asked about more randomly chosen villagers.

3 Standard errors clustered at the village level.

4 *** p<0.01, ** p<0.05, * p<0.1

Table 5: Diffusion Within the Village

| | Heard of PP | | | Knows how to PP | | | Adopts PP | | |
|---|---|---|---|---|---|---|---|---|---|
| | (1) | (2) | (3) | (4) | (5) | (6) | (7) | (8) | (9) |
| Connected to 1 seed | 0.002 | 0.030 | 0.016 | 0.017 | 0.021 | -0.031 | 0.008 | 0.012 | 0.004 |
| | (0.024) | (0.022) | (0.029) | (0.016) | (0.017) | (0.023) | (0.011) | (0.015) | (0.017) |
| Connected to 2 seeds | 0.084 ** | 0.124 *** | 0.064 | 0.062 ** | 0.068 ** | 0.110 ** | 0.016 | 0.039 ** | 0.014 |
| | (0.038) | (0.040) | (0.064) | (0.028) | (0.029) | (0.051) | (0.014) | (0.019) | (0.035) |
| Within path length 2 of at least one seed | -0.018 | 0.016 | 0.067 | 0.005 | 0.022 | 0.028 | 0.013 | 0.022 * | 0.037 * |
| | (0.028) | (0.027) | (0.042) | (0.018) | (0.021) | (0.028) | (0.008) | (0.013) | (0.021) |
| Season | 1 | 2 | 3 | 1 | 2 | 3 | 1 | 2 | 3 |
| N | 4155 | 4532 | 3103 | 4155 | 4532 | 3103 | 4203 | 3931 | 2998 |
| Mean of Reference Group (No connection to any seed) | 0.223 | 0.286 | 0.391 | 0.057 | 0.095 | 0.147 | 0.013 | 0.044 | 0.043 |
| SD of Reference Group | 0.416 | 0.452 | 0.488 | 0.232 | 0.293 | 0.355 | 0.113 | 0.206 | 0.203 |
| *p-value* for 2 connections = 1 connection | 0.018 | 0.013 | 0.442 | 0.072 | 0.091 | 0.004 | 0.522 | 0.164 | 0.760 |

Notes

1 Sample excludes seed and shadow farmers. Only connections to simple, complex and geo seed farmers are considered (no connections to benchmark farmers included).

2 The dependent variable in columns (1)-(3) is an indicator for whether the respondent reported being aware of a plot preparation method other than ridging and then subsequently indicated awareness of pit planting in particular. In columns (4)-(6), the dependent variable is an indicator for whether the farmer reported knowing how to implement pit planting. The dependent variable in (7)-(9) is an indicator for the household having adopted pit planting in that season.

3 In all columns, additional controls include indicators for the respondent being connected to: one Simple partner, two Simple partners, one Complex partner, two Complex partners, one Geo partner, two Geo partners, within 2 path length of a Simple partner, within 2 path length of a Complex Partner, and within 2 path length of the geo partner.

4 Also included in both panels are village fixed effects. Standard errors clustered at the village level.

5 The reference group is comprised of individuals with no direct or 2-path-length connections to a seed farmer.

6 *** $p<0.01$, ** $p<0.05$, * $p<0.1$

Table 6: Village-Level Regressions of Adoption Outcomes Across Treatment Arms

| | Any Non-Seed Adopters | | | | Adoption Rate | | | |
|---|---|---|---|---|---|---|---|---|
| | (1) | | (2) | | (3) | | (4) | |
| Simple Contagion Treatment | 0.155 | | 0.189 | * | 0.036 | ** | 0.006 | |
| | (0.100) | | (0.111) | | (0.017) | | (0.022) | |
| Complex Contagion Treatment | 0.252 | *** | 0.304 | *** | 0.036 | ** | 0.036 | |
| | (0.093) | | (0.101) | | (0.016) | | (0.026) | |
| Geographic treatment | 0.107 | | 0.188 | * | 0.038 | | 0.013 | |
| | (0.096) | | (0.110) | | (0.027) | | (0.034) | |
| Year | 2 | | 3 | | 2 | | 3 | |
| N | 200 | | 141 | | 200 | | 141 | |
| Mean of Benchmark Treatment (omitted category) | 0.420 | | 0.543 | | 0.038 | | 0.075 | |
| SD of Benchmark | 0.499 | | 0.505 | | 0.073 | | 0.109 | |
| *p-values for equality in coefficients:* | | | | | | | | |
| Simple = Complex | 0.300 | | 0.240 | | 0.981 | | 0.173 | |
| Complex = Geo | 0.102 | | 0.220 | | 0.937 | | 0.491 | |
| Simple = Geo | 0.623 | | 0.990 | | 0.950 | | 0.783 | |

Notes

1 The reference group is the Benchmark treatment.

2 The adoption rate in columns (1)-(2) include all radomly sampled farmers, excluding seed and shadow farmers. The "Any non-seed adopters" indicator in columns (3)-(4) excludes only seed farmers.

3 Sample for year 3 (columns 2 and 4) excludes Nkhotakota district.

4 All columns include controls used in the re-randomization routine (percent of village using compost at baseline; percent village using fertilizer at baseline; percent of village using pit planting at baseline); village size and its square; and district fixed effects. Standard errors are clustered at the village level.

5 *** p<0.01, ** p<0.05, * p<0.1

Table 7: Heterogeneity in Farmer-Level Adoption Decisions Across Treatment Arms

| | (1) | | (2) | | (3) | | (4) | |
|---|---|---|---|---|---|---|---|---|
| Bad Signal*simple | -0.008 | | -0.036 | | 0.019 | | -0.008 | |
| | (0.024) | | (0.037) | | (0.017) | | (0.034) | |
| Bad Signal* complex | 0.006 | | -0.027 | | 0.013 | | -0.045 | |
| | (0.024) | | (0.036) | | (0.015) | | (0.033) | |
| Bad Signal * geo | 0.002 | | -0.068 | ** | 0.031 | | -0.054 | * |
| | (0.031) | | (0.031) | | (0.035) | | (0.032) | |
| Good Signal | -0.037 | ** | -0.062 | ** | -0.007 | | -0.064 | * |
| | (0.017) | | (0.024) | | (0.022) | | (0.038) | |
| Good Signal * Simple | 0.064 | *** | 0.029 | | 0.054 | * | 0.021 | |
| | (0.021) | | (0.020) | | (0.029) | | (0.020) | |
| Good Signal * Complex | 0.059 | *** | 0.067 | *** | 0.054 | ** | 0.083 | *** |
| | (0.018) | | (0.025) | | (0.024) | | (0.030) | |
| Good Signal * Geo | 0.042 | ** | 0.022 | | 0.026 | | 0.031 | |
| | (0.020) | | (0.023) | | (0.022) | | (0.029) | |
| Good Signal Type | Flat Land | | Flat Land | | Unfamilliar Tech | | Unfamilliar Tech | |
| Year | 2 | | 3 | | 2 | | 3 | |
| N | 3546 | | 2645 | | 3954 | | 3023 | |
| Mean of Bad Signal in Benchmark Treatment | 0.066 | | 0.123 | | 0.046 | | 0.104 | |
| SD | 0.248 | | 0.33 | | 0.21 | | 0.305 | |
| *p-values for equality in coefficients:* | | | | | | | | |
| Simple, Good = Complex, Good | 0.828 | | 0.113 | | 0.986 | | 0.032 | |
| Complex, Good = Geo, Good | 0.482 | | 0.103 | | 0.297 | | 0.138 | |
| Simple, Good = Geo, Good | 0.364 | | 0.755 | | 0.351 | | 0.680 | |

Notes

1 The reference group is Bad signal recipients in the Benchmark treatment.

2 In columns (1)-(2), households with any flat land are those who have Good Signal=1 and those with all sloped land have Good Signal=0. In columns (3)-(4), households in villages where less than 4.32% of households (the median) ever tried pit planting at baseline are those who have Good Signal=1.

3 Sample for year 3 (columns 2 and 4) covers only 2 districts (Mwanza and Machinga).

4 All columns include controls used in the re-randomization routine (percent of village using compost at baseline; percent village using fertilizer at baseline; percent of village using pit planting at baseline); village size and its square; and district fixed effects. Standard errors are clustered at the village level.

5 *** p<0.01, ** p<0.05, * p<0.1

Figure A1: Project Timeline

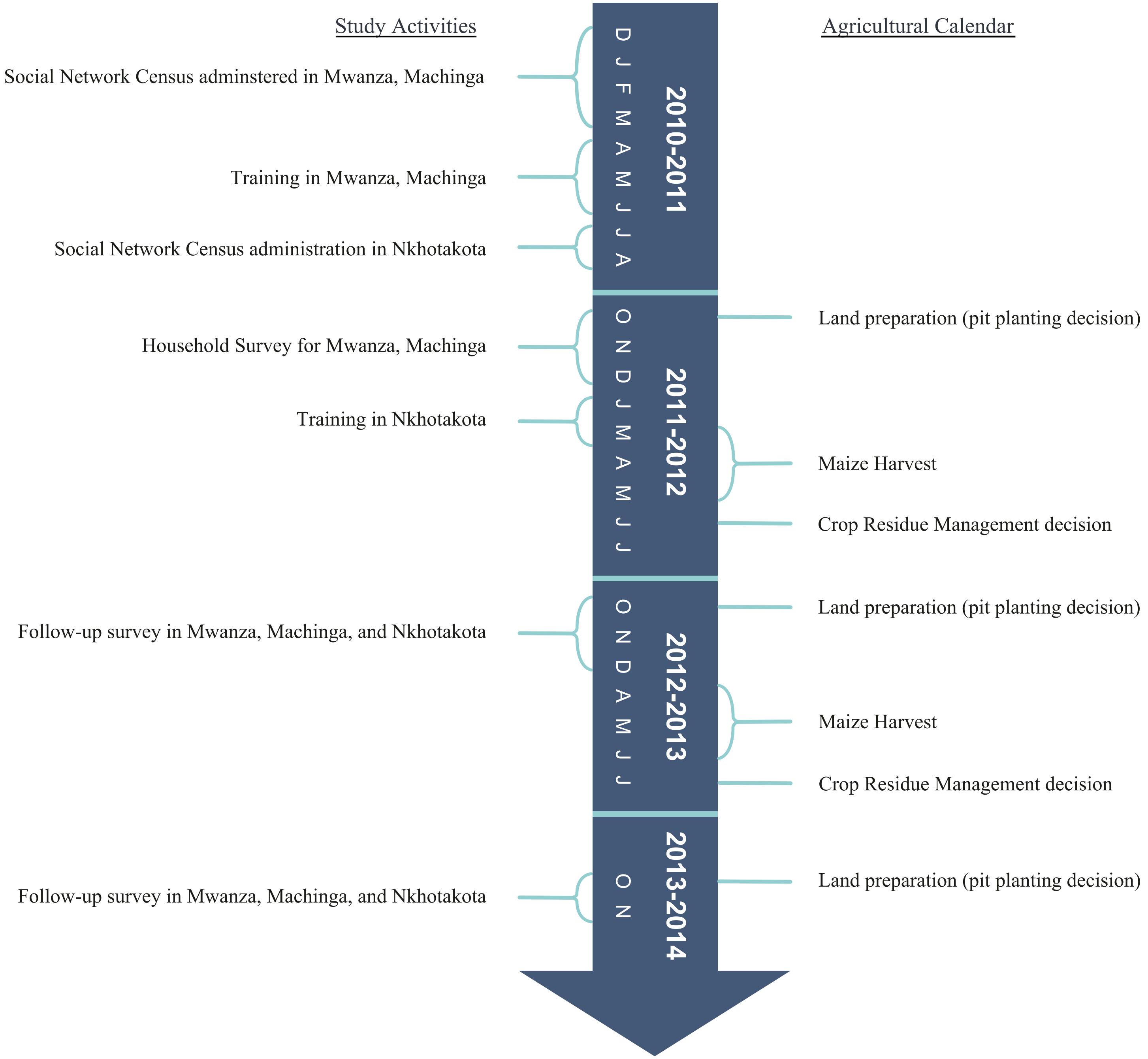

Table A1: Test of Balance across Randomized Treatment Arms

| | Housing | Assets | Livestock | Basal fertiliser (kg) | Top dressing fertiliser (kg) | # of Adults | # of Children | Farm size (acres) | Own land | Yields | Provided Ganyu | Used Ganyu |
|---|---|---|---|---|---|---|---|---|---|---|---|---|
| | (1) | (2) | (3) | (4) | (5) | (7) | (8) | (9) | (10) | (11) | (12) | (13) |
| Benchmark | 0.133 | 0.0327 | 0.0251 | 52.12 | 53.13 | 2.323 | 2.617 | 1.825 | 0.912 | 302.6 | 0.230 | 0.145 |
| | (0.031) | (0.027) | (0.028) | (2.699) | (2.154) | (0.016) | (0.033) | (0.040) | (0.005) | (5.469) | (0.007) | (0.006) |
| Simple Treatment | 0.142 | -0.035 | 0.023 | 52.57 | 51.53 | 2.313 | 2.625 | 1.628 | 0.905 | 307.3 | 0.252 | 0.129 |
| | (0.029) | (0.025) | (0.026) | (2.522) | (1.999) | (0.015) | (0.031) | (0.037) | (0.004) | (5.122) | (0.007) | (0.005) |
| Complex Treatment | -0.009 | -0.010 | 0.032 | 53.84 | 51.51 | 2.327 | 2.676 | 1.690 | 0.908 | 295.7 | 0.248 | 0.138 |
| | (0.032) | (0.027) | (0.028) | (2.732) | (2.179) | (0.016) | (0.034) | (0.040) | (0.005) | (5.547) | (0.007) | (0.006) |
| Geo Treatment | 0.027 | -0.023 | -0.082 | 51.60 | 50.97 | 2.309 | 2.643 | 1.771 | 0.905 | 304.3 | 0.244 | 0.151 |
| | (0.031) | (0.026) | (0.027) | (2.635) | (2.102) | (0.015) | (0.033) | (0.039) | (0.005) | (5.369) | (0.007) | (0.006) |
| Observations | 14,027 | 14,284 | 14,284 | 10,374 | 10,475 | 14,041 | 14,284 | 14,021 | 14,284 | 13,438 | 14,016 | 14,016 |
| *p-value for…* | | | | | | | | | | | | |
| Control = Simple | 0.004 | 0.131 | 0.457 | 0.998 | 0.441 | 0.965 | 0.335 | 0.011 | 0.838 | 0.915 | 0.215 | 0.033 |
| Control = Complex | 0.198 | 0.213 | 0.586 | 0.657 | 0.327 | 0.718 | 0.240 | 0.072 | 0.953 | 0.399 | 0.302 | 0.212 |
| Control = Geo | 0.548 | 0.327 | 0.065 | 0.901 | 0.629 | 0.980 | 0.519 | 0.618 | 0.812 | 0.811 | 0.759 | 0.575 |
| Simple = Complex | 0.150 | 0.696 | 0.776 | 0.708 | 0.897 | 0.613 | 0.664 | 0.554 | 0.856 | 0.389 | 0.944 | 0.280 |
| Simple = Geo | 0.325 | 0.624 | 0.273 | 0.925 | 0.705 | 0.989 | 0.802 | 0.124 | 0.924 | 0.871 | 0.496 | 0.026 |
| Complex = Geo | 0.808 | 0.875 | 0.105 | 0.573 | 0.570 | 0.711 | 0.542 | 0.322 | 0.840 | 0.480 | 0.562 | 0.121 |
| Joint | 0.034 | 0.446 | 0.210 | 0.949 | 0.775 | 0.959 | 0.662 | 0.058 | 0.993 | 0.810 | 0.598 | 0.076 |

Notes

1 Housing, assets and livestock are pca scores. Housing includes information on: materials walls are made of, roof materials, floor materials and whether the household has a toilet. Assets includes the number of bycicles, radios and cell phones the household owns. Livestock is an index including the number of sheep, goats, chickens, cows, pigs guinea fowl, and doves.

2 Standard errors are in parentheses.

3 *** p<0.01, ** p<0.05, * p<0.1

Table A2: Frequency with which Different Targeting Strategies Select the same Individuals as Seeds

| | Complex | Geo | Benchmark |
|---|---|---|---|
| Simple | 25.51% | 3.83% | 10.11% |
| Complex | | 6.00% | 12.36% |
| Geo | | | 6.74% |

Notes

1 Each row indicates the probability that a potential seed of the targeting strategy indicated in the row is also a seed of the targeting strategy indicated in the column.

2 The Benchmark column examines only villages assigned to the benchmark treatment. As an example, 10.11% of seeds in Benchmark villages are also potential Simple seeds.

Table A3: Simulation of Village Level Adoption Outcomes across all treatment cells, assuming Diffusion follows either Complex or Simple Contagion Pattern

| | Simulated Adoption Rate | | | | Simulated Any Adopters | | | |
|---|---|---|---|---|---|---|---|---|
| | (1) | | (2) | | (3) | | (4) | |
| **Panel A: Simulations Assuming Farmers learn by Simple Contagion** | | | | | | | | |
| Simple Treatment | 0.026 | | 0.090 | * | 0.095 | ** | 0.036 | |
| | (0.024) | | (0.052) | | (0.043) | | (0.037) | |
| Complex Treatment | 0.087 | *** | 0.072 | | 0.060 | | 0.013 | |
| | (0.029) | | (0.063) | | (0.048) | | (0.045) | |
| Geo treatment | -0.022 | | -0.113 | ** | -0.050 | | -0.070 | |
| | (0.027) | | (0.057) | | (0.053) | | (0.054) | |
| Year | 2 | | 3 | | 2 | | 3 | |
| N | 187 | | 138 | | 187 | | 138 | |
| Mean Benchmark Partners | 0.182 | | 0.504 | | 0.845 | | 0.927 | |
| SD Benchmark Partners | 0.149 | | 0.306 | | 0.258 | | 0.186 | |
| Test: Simple = Complex | 0.013 | | 0.733 | | 0.384 | | 0.559 | |
| Test: Complex = Geo | 0.000 | | 0.001 | | 0.026 | | 0.129 | |
| Test: Simple = Geo | 0.035 | | 0.000 | | 0.001 | | 0.030 | |
| **Panel B: Simulations Assuming Farmers Learn by Complex Contagion** | | | | | | | | |
| Simple Treatment | 0.001 | | -0.022 | | -0.092 | | -0.109 | |
| | (0.012) | | (0.040) | | (0.056) | | (0.077) | |
| Complex Treatment | 0.047 | *** | 0.162 | *** | 0.257 | *** | 0.275 | *** |
| | (0.012) | | (0.046) | | (0.061) | | (0.081) | |
| Geo treatment | 0.008 | | -0.032 | | -0.028 | | -0.048 | |
| | (0.011) | | (0.038) | | (0.060) | | (0.083) | |
| Season | 2 | | 3 | | 2 | | 3 | |
| N | 187 | | 138 | | 187 | | 138 | |
| Mean Benchmark Partners | 0.038 | | 0.138 | | 0.436 | | 0.541 | |
| SD Benchmark Partners | 0.0479 | | 0.194 | | 0.341 | | 0.39 | |
| Test: Simple = Complex | 0.000 | | 0.000 | | 0.000 | | 0.000 | |
| Test: Complex = Geo | 0.001 | | 0.000 | | 0.000 | | 0.000 | |
| Test: Simple = Geo | 0.533 | | 0.777 | | 0.192 | | 0.370 | |

Notes

1 Simulations only include control villages where we had both seeds in social network census.

Table A4: Individual-level analysis of Crop Residue Management Familiarity and Adoption

| | Season 1 | | | | Season 2 | | | |
|---|---|---|---|---|---|---|---|---|
| | (1) | | (2) | | (3) | | (4) | |
| | Adopted CRM | | Aware of CRM | | Adopted CRM | | Aware of CRM | |
| Connected to one seed | -0.026 | | 0.045 | ** | 0.005 | | -0.042 | * |
| | (0.020) | | (0.022) | | (0.028) | | (0.023) | |
| Connections to two seeds | -0.011 | | 0.096 | ** | 0.057 | | -0.021 | |
| | (0.034) | | (0.037) | | (0.047) | | (0.038) | |
| N | 4203 | | 4149 | | 2683 | | 4531 | |
| Mean of Excluded Group | 0.239 | | 0.295 | | 0.159 | | 0.366 | |
| SD of Excluded Group | 0.427 | | 0.456 | | 0.366 | | 0.482 | |
| *p* -value of Test: 2 connections = 1 connection | 0.661 | | 0.143 | | 0.189 | | 0.593 | |

Notes

1 Sample excludes seed and shadow farmers in all villages, and excludes control villages. Seed farmers are either simple, control or geo (no benchmark farmers included).

2 Additional controls include indicators for the respondent being connected to: one Simple partner, two Simple partners, one Complex partner, two Complex partners, one Geo partner and two Geo partners.

3 Also included in both panels are village fixed effects.

4 The reference group is comprised of individuals with no connections to a seed farmer.

5 Columns 1, 2 and 4 include data on Mwanza, Machinga and Nkhotakota. Column 3 includes only Mwanza and Machinga, as the final round of data collection in Nkhotakota was prior to the decision on whether to use CRM.